\documentclass[jgrga]{agutex}

\usepackage{color}

  \usepackage{graphicx, epstopdf}    
  \setkeys{Gin}{draft=false}
\authorrunninghead{LOI ET AL.}

\titlerunninghead{DENSITY DUCT FORMATION: MWA OBSERVATIONS}

\authoraddr{Corresponding author: S.~T.~Loi,
Sydney Institute for Astronomy, School of Physics, University of Sydney, NSW 2006, Australia. (stl36@cam.ac.uk)}

\begin{document}

%
%

\title{Density duct formation in the wake of a travelling ionospheric disturbance: Murchison Widefield Array observations}
%
%

%
%









\authors{
Shyeh Tjing Loi\altaffilmark{1,2},
Iver H.~Cairns\altaffilmark{1},
Tara Murphy\altaffilmark{1,2},
Philip J.~Erickson\altaffilmark{3},
Martin E.~Bell\altaffilmark{2,4},
Antonia Rowlinson\altaffilmark{2,4},
Balwinder Singh Arora\altaffilmark{5},
John Morgan\altaffilmark{5},
Ronald D.~Ekers\altaffilmark{4},
Natasha Hurley-Walker\altaffilmark{5}, 
David L.~Kaplan\altaffilmark{6}
}

\altaffiltext{1}{Sydney Institute for Astronomy, School of Physics, University of Sydney, Sydney, New South Wales, Australia}

\altaffiltext{2}{ARC Centre of Excellence for All-sky Astrophysics (CAASTRO)}

\altaffiltext{3}{MIT Haystack Observatory, Westford, Massachusetts, USA}

\altaffiltext{4}{CSIRO Astronomy and Space Science, Epping, New South Wales, Australia}

\altaffiltext{5}{International Centre for Radio Astronomy Research, Curtin University, Bentley, Western Australia, Australia}

\altaffiltext{6}{Department of Physics, University of Wisconsin-Milwaukee, Milwaukee, Wisconsin, USA}

%
%


\begin{abstract}
Geomagnetically-aligned density structures with a range of sizes exist in the near-Earth plasma environment, including 10--100\,km-wide VLF/HF wave-ducting structures. Their small diameters and modest density enhancements make them difficult to observe, and there is limited evidence for any of the several formation mechanisms proposed to date. We present a case study of an event on 26 August 2014 where a travelling ionospheric disturbance (TID) shortly precedes the formation of a complex collection of field-aligned ducts, using data obtained by the Murchison Widefield Array (MWA) radio telescope. Their spatiotemporal proximity leads us to suggest a causal interpretation. Geomagnetic conditions were quiet at the time, and no obvious triggers were noted. Growth of the structures proceeds rapidly, within 0.5\,hr of the passage of the TID, attaining their peak prominence 1--2\,hr later and persisting for several more hours until observations ended at local dawn. Analyses of the next two days show field-aligned structures to be preferentially detectable under quiet rather than active geomagnetic conditions. We used a raster scanning strategy facilitated by the speed of electronic beamforming to expand the quasi-instantaneous field of view of the MWA by a factor of three. These observations represent the broadest angular coverage of the ionosphere by a radio telescope to date.
\end{abstract}

%
%

%

\begin{article}

%
%

\section{Introduction}\label{sec:intro}
\subsection{Physical Background}\label{sec:intro_phys}
Radio propagation is powerful probe of the density distribution of magnetospheric and ionospheric plasma. The refractive, diffractive and dispersive effects that free electrons have on radio signals enable structure to be deduced from whistler spectrograms \citep{Sagredo1973, Singh1994}, observations of cosmic radio sources \citep{Wild1956, Coker2009, Helmboldt2014, Loi2015_mn2e}, sounding and radar scattering experiments \citep{Hunsucker1991}, and satellite dual-frequency delay measurements \citep{Mannucci1998, Bernhardt2006, Hawarey2006}. The geomagnetic field imposes significant anisotropy in the shapes of density irregularities, since electron and ion mobilities are much larger parallel than perpendicular to the magnetic field \citep{Calvert1969, Booker1979}. Consequently, irregularities over a huge range of scales, from metres to thousands of kilometres \citep{Park1970, Fejer1980, Darrouzet2009}, tend to be elongated along the field. Their axial ratios can be very large, since in extreme cases the structures bridge field lines between conjugate hemispheres to form ducts capable of guiding VLF to HF radio signals \citep{Storey1953, Smith1960, Park1971, Platt1989}. Other means of identifying field-aligned ducts include satellite in-situ measurements \citep{Angerami1970, Sonwalkar1994} and radio interferometric observations \citep{Jacobson1993, Hoogeveen1997a, Helmboldt2012b, Loi2015_mn2e}. 

Properties deduced from observations and modelling \citep{Park1970, Sagredo1973, Walker1978, Lester1980} are cross-sectional widths of 10--100\,km, enhancements and depletions of 1--10\% relative to the ambient density, and timescales of growth and decay of hours to days. Whistler-mode waves, which can be trapped and guided by these ducts, are believed to play a major role in the precipitation of energetic particles through resonant interactions \citep{Voss1984, Burgess1993, Abel1998}. By confining the wave normals along the magnetic field and the wave fields within the ducts, these ducts greatly enhance the interaction efficiency between particles and natural or human generated VLF waves. Suggestions are that they could be important for regulating the overall equilibrium level of the radiation belts \citep{James2006, Sonwalkar2006, Kulkarni2008}. A better knowledge of their occurrence rates would improve global models of magnetospheric-ionospheric coupling. More broadly, there is practical relevance for understanding their contribution to signal distortion in radio communications and radio astronomy, so that suitable mitigation strategies may be developed.

Early work discussed the idea of a convective instability where the presence of an insulating layer (the neutral atmosphere) allows material in magnetic flux tubes to be interchanged in a manner that is symmetric between conjugate hemispheres of the Earth \citep{Gold1959, Newcomb1961}. This is one example of a process that can give rise to field-aligned ducts. It is difficult to ascertain how these form in reality \citep{Singh1998, Darrouzet2009}, and many triggers have been proposed. These include electric field fluctuations arising from turbulence in the ionospheric dynamo region \citep{Dagg1957a} and the outer magnetosphere \citep{Reid1965}, conductivity irregularities in the lower ionosphere \citep{Cole1971}, thundercloud electric fields \citep{Park1971}, and irregularities in the ring current particle density where it overlaps the plasmasphere \citep{Walker1978}. While these differ in the source of electric field, they all invoke the same physical process, namely the interchange of magnetic flux tubes under $\mathbf{E}\times\mathbf{B}$ drift, which generates fine-scale density structure by permuting the positions of flux tubes with differing electron content. 

The thundercloud mechanism of \citet{Park1971} has received the most attention thus far \citep[e.g.][]{Rodger1998}, but its viability remains controversial. While supported by theoretical modelling \citep{Park1973, Rodger2002, McCormick2002}, observations find no compelling link between duct numbers and thunderstorm activity \citep{Clilverd2001}. The ring current overlap mechanism of \citet{Walker1978}, which predicts a most probable duct formation region of 1.5 $R_\oplus$ (where $R_\oplus$ is an Earth radius) within the plasmapause during magnetically disturbed conditions, is consistent with measured increases in duct numbers during storm recovery periods \citep{Smith1991}. Multiple mechanisms may even operate simultaneously over the globe at any one time.

While density ducts are largely stationary with respect to the background plasma, density variations are also associated with moving disturbances. Acoustic-gravity waves (AGWs) propagate through the neutral atmosphere at speeds in excess of about 100\,m\,s$^{-1}$, the restoring force being a combination of gas pressure and gravity \citep{Hines1960, Georges1968}. AGWs are produced by various meteorological, seismic and geomagnetic events \citep{Hunsucker1982, Kazimirovsky2002, Liu2011}. Those generated at low altitudes (below $\sim$200\,km), for example by hurricanes or nuclear explosions, refract upwards to F-region heights under natural temperature gradients where they drive modulations in the electron density \citep{Richmond1978}. The density fluctuations associated with AGWs have also been suggested as a source of the electric field needed for the development of field-aligned ducts \citep{Cole1971}.

\subsection{Technical Background}\label{sec:intro_tech}
Radio interferometers, which are often used to study celestial radio sources, can be used to probe the density distribution of the near-Earth plasma. Spatial variations in the electron column density (total electron content, TEC) produce phase shifts of received signals. The high precision to which phases may be measured allows radio interferometers to probe TEC variations at high sensitivity: for instance, the Very Large Array (VLA) can achieve a precision of $\sim$0.1\,mTECU\,km$^{-1}$ (1\,TECU = $10^{16}$\,el\,m$^{-2}$) at 74\,MHz \citep{Helmboldt2012d}. In practice, this may be achieved either by measuring the apparent angular fluctuations of radio sources with known sky positions \citep{Meyer-Vernet1980, Bougeret1981, Mercier1986, Helmboldt2012b, Helmboldt2012c, Cohen2009, Loi2015_mn2e} or by observing a strong source and comparing the phase fluctuations between receivers \citep{Jacobson1992, Jacobson1992a, Hoogeveen1997, Hoogeveen1997a, Dymond2011_mn2e, Helmboldt2012, Helmboldt2014a}. Information about smaller-scale structure comes from analysing the scintillation of point-like sources with fixed intrinsic brightness \citep{Hewish1951, Wild1956, Spoelstra1985, Narayan1989, Spoelstra1997}. The distribution of sampling points is given by the configuration of sight lines from receivers to background sources, and the sampling density by the number of receiving elements and/or the number of observable sources, depending on the method used.

\begin{figure}[H]
  \centering
  \includegraphics[width=\columnwidth]{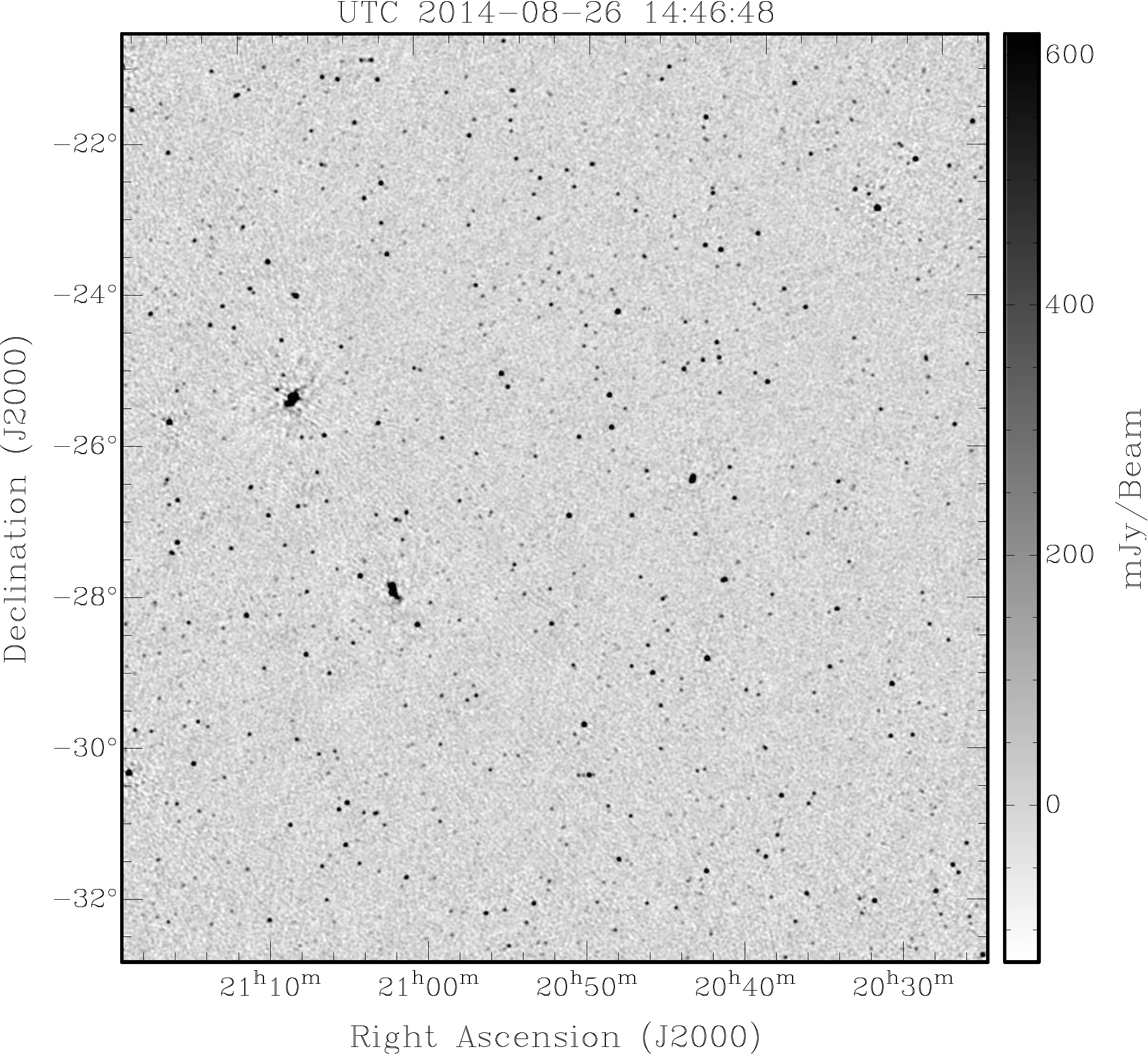}
  \caption{An example snapshot image integrated over 2\,min at 154\,MHz, zoomed in to the central quarter so that individual sources (mostly radio galaxies and quasars, appearing as unresolved point-like objects) are visible. The full image is around 1000\,deg$^2$ in size and contains $\sim$3300 sources above 5$\sigma$, where $\sigma$ is the pixel-to-pixel root-mean-square (RMS) noise.}
  \label{fig:2014Aug-26_snapshot}
\end{figure}

A new generation of low-frequency radio interferometers with wide fields of view (FoV) and large numbers $N$ of receiving elements is now emerging, with geospace monitoring capabilities that surpass the old generation of interferometers in terms of the breadth and detail they achieve. These include the Murchison Widefield Array \citep[MWA;][]{Lonsdale2009_mn2e, Bowman2013_mn2e, Tingay2013_mn2e} and the Low Frequency Array \citep[LOFAR;][]{vanHaarlem2013}. Recent work using the MWA to map the regional TEC distribution \citep{Loi2015_mn2e, Loi2015a_mn2e} has demonstrated the success of the wide-FoV, large-$N$ design of radio interferometer for surveying the near-Earth plasma \citep{Coster2012}.

\subsection{This Work}
We present a case study of an event that shows the formation of field-aligned density structures shortly after the passage of a travelling ionospheric disturbance (TID). We also analyse intervals from the subsequent two nights, the first geomagnetically active and the second quieter. For this work we have processed the data for the main event in such a way as to achieve a near-instantaneous FoV that is three times larger than the usual size of the MWA FoV, representing the broadest angular observation of the ionosphere by a radio telescope to date.

This paper is structured as follows. In Section \ref{sec:data} we describe the observing strategy and conditions under which the MWA data were obtained. In Section \ref{sec:method} we describe the data reduction/imaging process and analysis methods. We interpret and discuss the results in Section \ref{sec:results}, and conclude in Section \ref{sec:conclusion}.

\section{Observations}\label{sec:data}

\begin{figure}[H]
  \centering
  \includegraphics[width=\columnwidth]{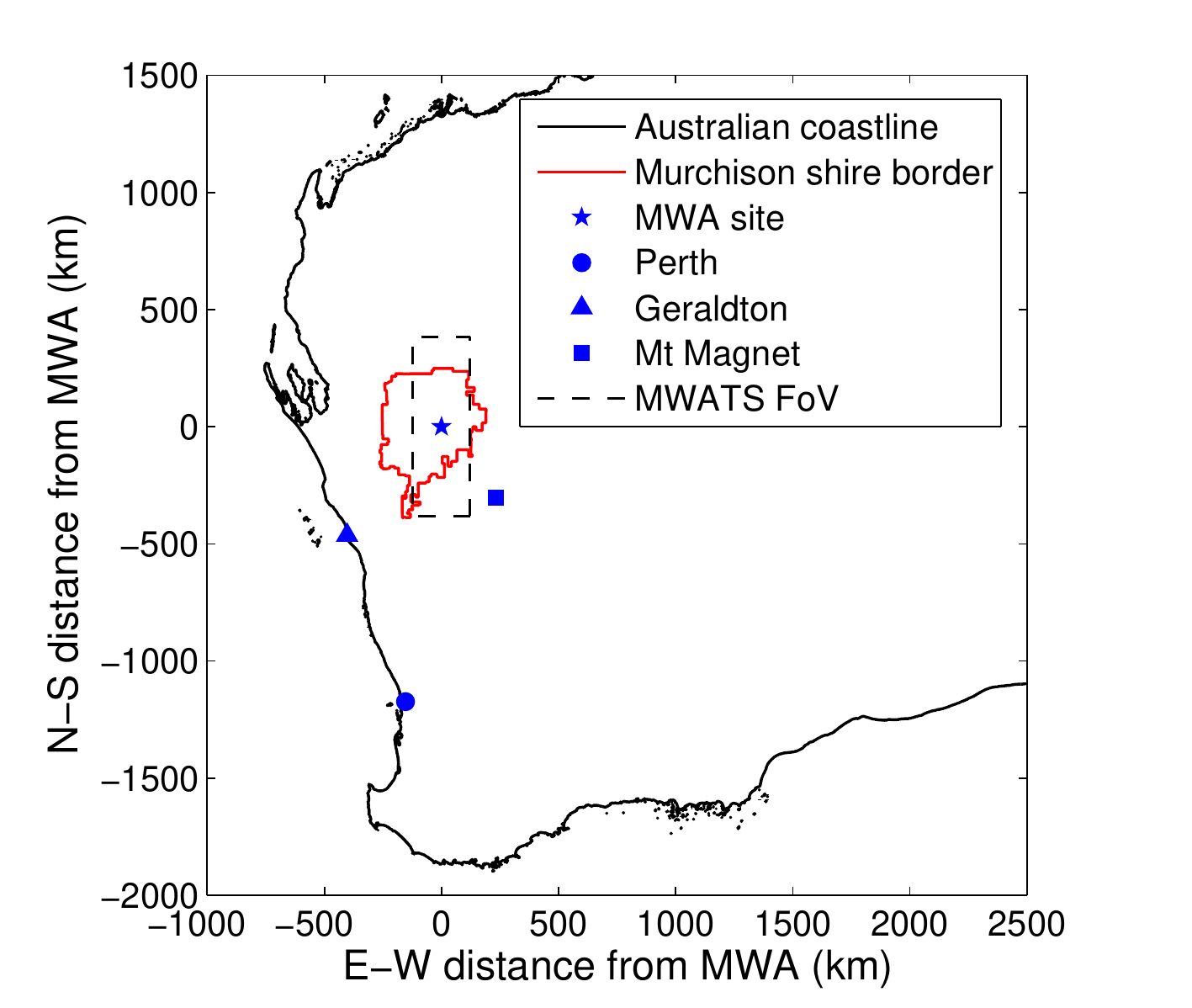}
  \caption{An illustration of the relevant geographical scales for MWA observations. The MWA is marked by the pentagram at the origin, with the Murchison shire border indicated by the red line (the shire is roughly the size of the Netherlands) and nearby towns marked by various symbols. The effective FoV spanned by the triple-pointing survey mode is given by the dashed rectangle, which was computed as the projection of the angular FoV onto an altitude of 350\,km.}
  \label{fig:geography}
\end{figure}

\subsection{MWA Instrument and Data}\label{sec:data_collection}
The MWA is located in the Murchison Shire of Western Australia at $26^\circ 42' 12''$S, $116^\circ 40' 15''$E (geographic) and a geomagnetic latitude of 38.6$^\circ$S ($L = 1.6$ \citep{McIlwain1961}). It operates between 80--300\,MHz and currently comprises 128 receiver ``tiles'', each with 16 dual-polarisation bowtie dipole antennas arranged in a $4 \times 4$ configuration. An analogue beamformer steers each tile beam, which can be re-pointed in a different direction almost instantaneously. The omnidirectionality of the MWA dipoles gives it a much larger FoV compared to traditional dish-based radio telescopes. The FWHM of the primary beam is $\sim$30$^\circ$ at 154\,MHz, which subtends 180\,km at 350\,km altitude. Being a phased array, it can also be re-pointed much more rapidly than mechanically steered dishes. Combining these features with its ability for high-cadence imaging, the MWA can achieve extremely fast survey speeds. Note, however, that multibeam observations are not possible for the current design of the MWA beamformers. Figure \ref{fig:2014Aug-26_snapshot} shows an example MWA image integrated over 2\,min of time and 30.72\,MHz of bandwidth centred at 154\,MHz.

The 128 tiles are arranged in a compact, pseudo-random configuration with the longest baseline around 3\,km in length. Given that this is much smaller than the physical distance subtended by the FoV at F-region heights, and that a large number ($\sim 10^3$) of point-like radio sources are visible within a typical snapshot, the TEC distribution can be obtained by measuring the apparent displacements of sources in images synthesised by the whole array (regime 3 of \citet{Lonsdale2005}). This dense collection of sight lines lets the MWA probe structure down to scales of $\sim 1^\circ$, which subtends a physical distance of $\sim$6\,km at 350\,km altitude. The angular displacement $\Delta \theta$ of a radio source (in radians) is related to the observing frequency $\nu$ (in Hz) and the TEC gradient transverse to the line of sight $\nabla_\perp$TEC (in el\,m$^{-3}$) by \citep{Smith1952}
\begin{equation}
  \Delta \theta = -\frac{40.3}{\nu^2} \nabla_\perp \mathrm{TEC} \:, \label{eq:posoffset}
\end{equation}
which at 154\,MHz can be measured to a precision of $\sim$10\,arcsec for a source at the sensitivity limit, corresponding to a $\nabla_\perp$TEC precision of the order 1\,mTECU\,km$^{-1}$. The negative sign indicates that displacements are toward the direction of decreasing TEC. Time-dependent instrumental phase fluctuations (e.g.~from temperature-induced changes in cable lengths) could be an independent and alternative source of angular shifts. However, due to the excellent phase stability of the MWA \citep{Hurley-Walker2014}, the associated $\nabla_\perp$TEC errors would only be $\sim$0.01\,mTEC\,km$^{-1}$ and so negligible compared to the ionospheric contribution.

The MWA collects data mostly for astronomical purposes. Avoidance of the Sun (whose instrumental response can be difficult to treat during imaging) and large dawn/dusk TEC gradients (which degrade astronomical observations) restricts most MWA operations to local night. At the time of year during which the data used in this study were obtained (August), this corresponds to data collection between around 7\,pm and 5\,am local time (1100--2100 UT). We consider the three-day period between 26--28 August 2014, analysing data from two astronomical surveys that ran during that period: the MWA Transients Survey (MWATS; at 154\,MHz, between 1100--2050 UT on the 26th) and the Epoch of Reionisation Transients Survey (EoRTS; at 183\,MHz, between 1450--2210 UT on the 27th and 28th). The instantaneous bandwidth was 30.72\,MHz throughout. MWATS uses a ``drift-scan'' strategy where the instrument is cycled in a raster fashion between three pointings fixed in Az/El and spaced by 30$^\circ$, lingering at each for 2\,min (details in \citet{Loi2015b_mn2e}). In contrast, EoRTS is configured to track a given patch of celestial sky at a time, and so the FoV slowly pans across the ionosphere as the Earth rotates. The celestial coordinates tracked on the 27th and 28th were RA/Dec = $0^\circ,-27^\circ$ for the first half of each night, and RA/Dec = $60^\circ,-27^\circ$ for the second half.

The main event of interest occurs during the (local) night of the 26th between about 1530--2050 UT, where a large-amplitude TID precedes the formation of regional-scale field-aligned structures. The subsequent two nights were analysed as a follow-up. The raster scanning stategy of MWATS expands the effective FoV to cover a roughly $90^\circ \times 30^\circ$ (N-S $\times$ E-W) region of the ionosphere, shown plotted in Figure \ref{fig:geography} for a projected height of 350\,km (typical height of the F region density maximum). This spans a sizeable fraction of the dimensions of the Australian continent at these heights ($\sim$20\% of the N-S extent), encompassing an area of about 100,000\,km$^2$.

\subsection{Geomagnetic Conditions}\label{sec:data_conditions}
\begin{figure*}[H]
  \centering
  \includegraphics[clip=true, trim=2cm 0cm 2cm 0cm, width=\textwidth]{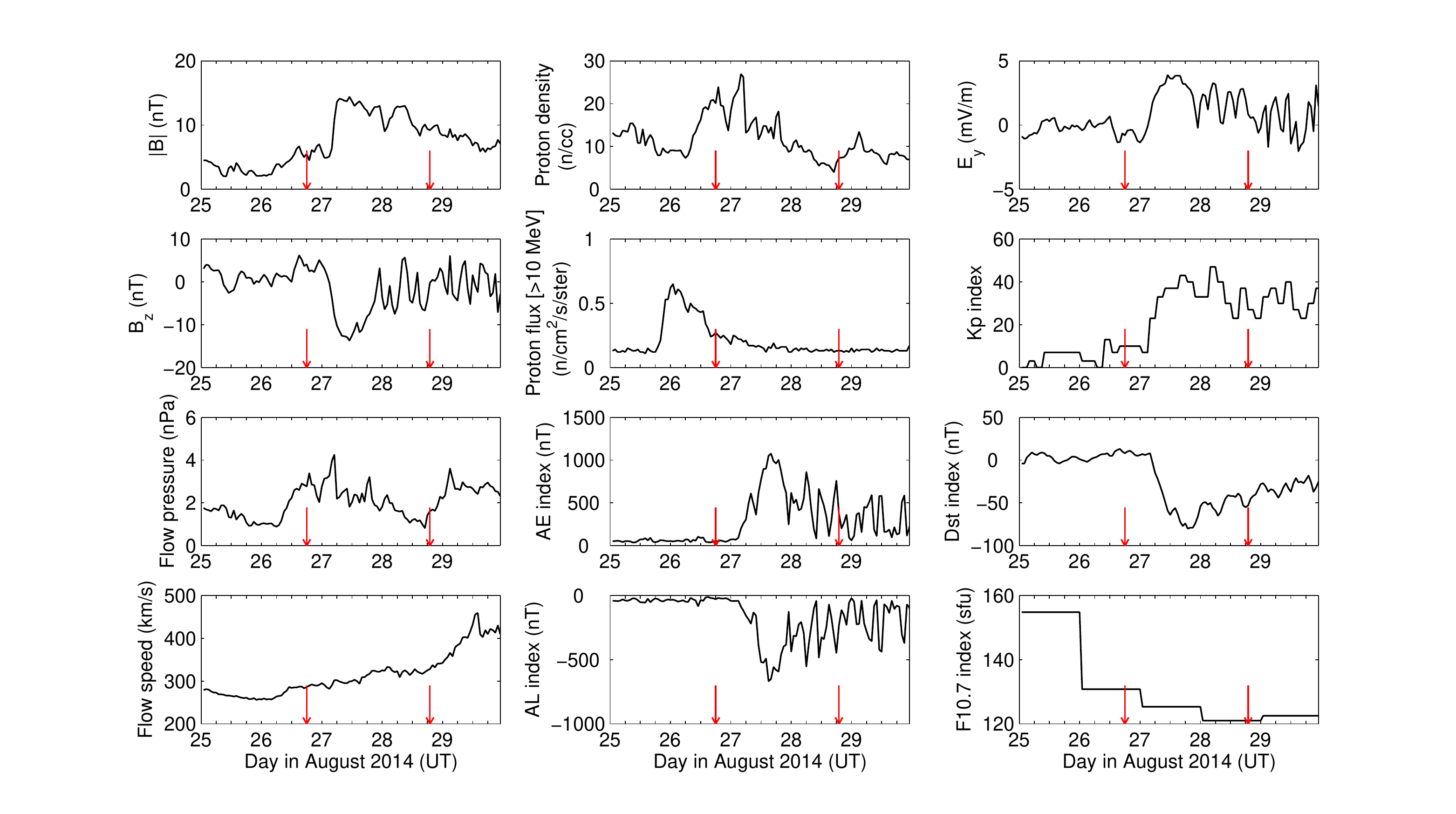}
  \caption{UT time dependence of space weather activity indices over the three-day period in question (26th to 28th August 2014), showing data for an additional preceding and succeeding day. Data are hourly averages obtained from the NASA OMNIWeb database. Conditions are mostly quiet for the first day analysed (26th) although an abrupt rise is noted in the energetic proton flux near 00 UT. A geomagnetic storm begins early on the 27th, and conditions remain moderately disturbed through the 28th. MWA data collection is not continuous but takes place exclusively at local night, with observation blocks usually occupying the second halves of each UT day. The times at which the TIDs discussed in Sections \ref{sec:results_26th} and \ref{sec:results_28th} appear are marked by red arrows.}
  \label{fig:omni}
\end{figure*}

Plots of the various disturbance indicators for the interval in question are shown in Figure \ref{fig:omni}. Geomagnetic conditions were very quiet on and leading up to the 26th. The most recent excursion in $D_{st}$ (the disturbance storm-time index, quantifying the strength of the ring current) \citep{Sugiura1964} took place on the 21st and fell only to $-30$\,nT, corresponding to a weak disturbance. The solar wind speed was low in the preceding days, being below 300\,km\,s$^{-1}$ onwards from the 24th and reaching a minimum of about 250\,km\,s$^{-1}$ early (in UT) on the 26th. A spike in energetic proton flux was noted to occur early on the 26th, but this was not accompanied by any excursions in $K_p$ (the planetary storm index), $D_{st}$, or the auroral electrojet indices. 

Conditions became much more disturbed early on the 27th after the arrival of a pair of slow-moving coronal mass ejections (CMEs) launched on the 22nd, marking the start of a moderate geomagnetic storm. Although the CMEs were moving too slowly to produce positive excursions in $D_{st}$ or the ambient flow speed, their geoeffectiveness was enhanced by their carrying a negative value of $B_z$ (southwards orientation of the magnetic field). The $K_p$ index reached 4 near the middle of the UT day, and the $D_{st}$ index reached a minimum of $-80$\,nT late on the 27th, completing the main storm phase period. Disturbed conditions continued through the 28th, with the recovery phase punctuated by a series of substorm events indicated by large negative spikes in the AL (amplitude lower) index.

\section{Methods}\label{sec:method}

\subsection{Data Reduction and Imaging}
The refractive analysis described in the next subsection required the total-intensity radio images as input, which were first generated from the raw data by a Fourier inversion of the interferometric visibilities. Much of the raw data processing for this work made use of custom software developed for use with the MWA. These include the algorithm \textsc{aoflagger} for removing radio frequency interference \citep{Offringa2015} and the imaging algorithm \textsc{wsclean} \citep{Offringa2014}. Details of the reduction process for data taken on the 26th and 28th can be found in \citet{Loi2015b_mn2e} and \citet{Rowlinson2015_mn2e} respectively, which describe the procedure for generating the final total-intensity images that we analyse in the current work. The respective imaging cadences for the 26th, 27th and 28th were 6\,min, 2\,min and 30\,s. Data from the 27th were reduced using the same pipeline as for the 26th, with Hercules A as the calibrator and final images integrated over 2-min blocks. Note that data from the 28th were imaged in 30-s blocks and so have fourfold greater time resolution than the 27th, which in turn has threefold greater time resolution than the data on the 26th since this is imposed by the cyclic 6-min survey strategy for MWATS.

\begin{figure*}[H]
  \centering
  \includegraphics[width=0.49\textwidth]{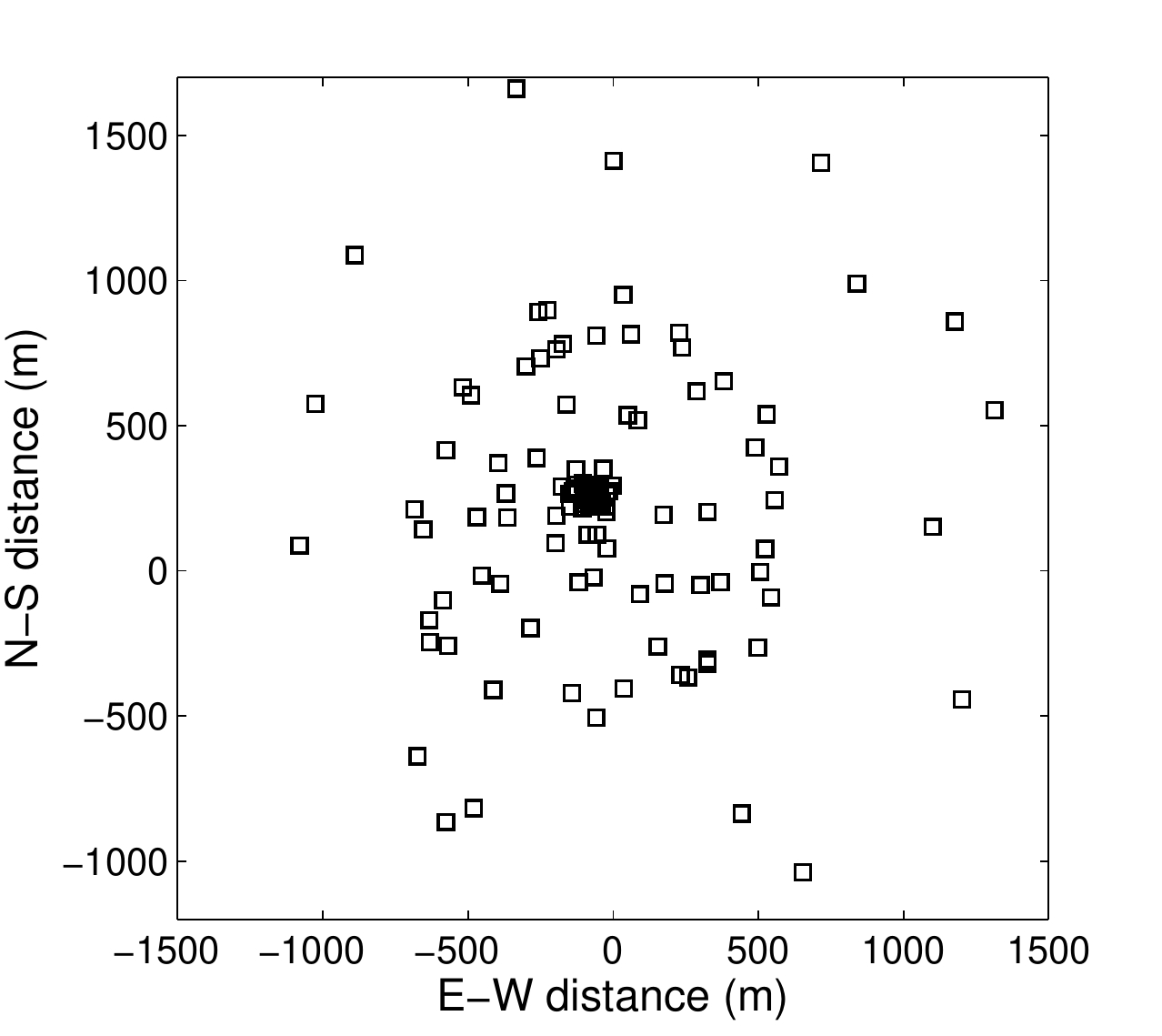}
  \includegraphics[width=0.49\textwidth]{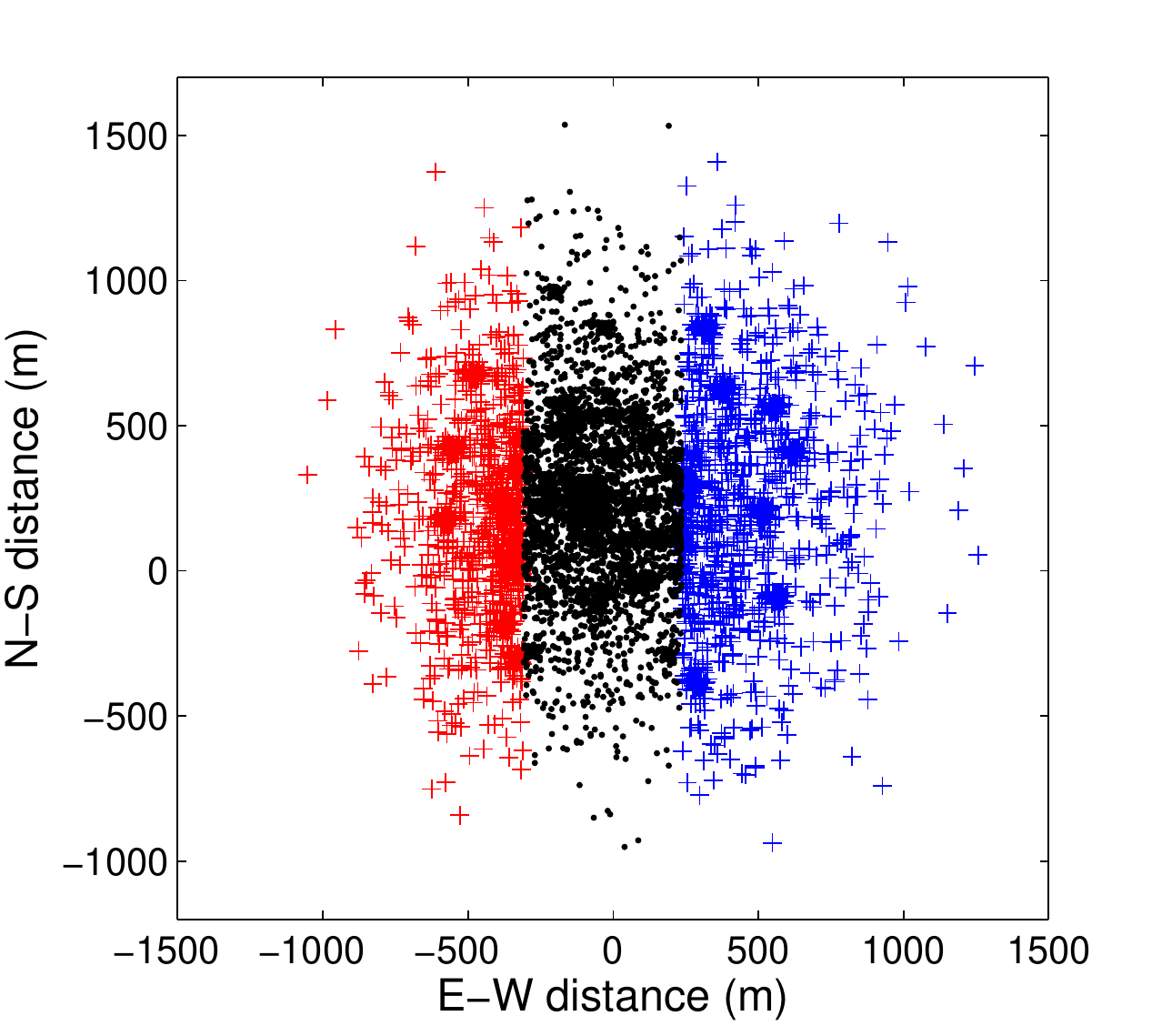}
  \caption{The spatial distribution of the 128 MWA antenna tiles (left panel) and the corresponding distribution of the $^{128}C_2 = 8128$ baseline midpoints (right panel), shown on identical axes. The coordinates are in terms of ground distance with east to the right and north up. Baselines used for parallax analysis are indicated by the `+' symbols, and excluded baselines by black dots. The east and west groupings of baselines are distinguished by the blue and red colours, respectively, and the separation of their centroids (the effective parallax baseline) is 880\,m.}
  \label{fig:mwalayout}
\end{figure*}

\subsection{Measuring the TEC}
Measuring the refractive shifts of known point-like cosmic radio sources from their reference positions allowed us to obtain a value for $\nabla_\perp$TEC using Equation (\ref{eq:posoffset}) in the direction of each source. The source finder software \textsc{Aegean} \citep{Hancock2012} was first used to identify candidate sources as local maxima in the images. These were cross-matched with published databases of known radio sources, retaining only those with counterparts. We used the National Radio Astronomy Observatory VLA Sky Survey catalogue \citep{Condon1998} for declinations northward of $-40^\circ$, and the Sydney University Molonglo Sky Survey catalogue \citep{Mauch2003} for declinations south of $-30^\circ$. The cross-matching radius was chosen to be 1.2\,arcmin for all data from EoRTS and the northern MWATS pointing, a value which was a reasonable tradeoff between false associations and the loss of highly displaced sources. However for the zenith and southern MWATS pointings, sources had undergone such large refractive displacements that we had to increase the cross-matching radius to 3.6\,arcmin to obtain a cross-matching efficiency (fraction of \textsc{Aegean} sources with a catalogue match) comparable to the other data ($\sim$95\%). 

For each appearance of a source, its refractive displacement was computed as the difference between its measured position and its time-averaged position, which we took to be the reference point. Repeating this measurement for all sources present allowed us to determine the $\nabla_\perp$TEC vector field, using Equation (\ref{eq:posoffset}). The angular TEC distribution is the two-dimensional spatial integral of $\nabla_\perp$TEC. Note that only the relative and not the absolute TEC value can be obtained, because a refractive analysis is insensitive to the constant offset component of the TEC (given the gradient of a function, that function can only be determined up to an additive constant). Our method for integrating $\nabla_\perp$TEC is described further in Appendix \ref{sec:relTEC}. Note that the method uses time-averaged rather than catalogue positions as the reference, thereby removing information about long-timescale fluctuations. This was done to avoid systematic errors resulting from imperfect phase calibration, which we previously found to produce global shifts in MWATS data comparable to ionospheric shifts (see section 2.3 of \citet{Loi2015a_mn2e} for details).

\begin{figure*}[H]
  \centering
  \includegraphics[width=0.49\textwidth]{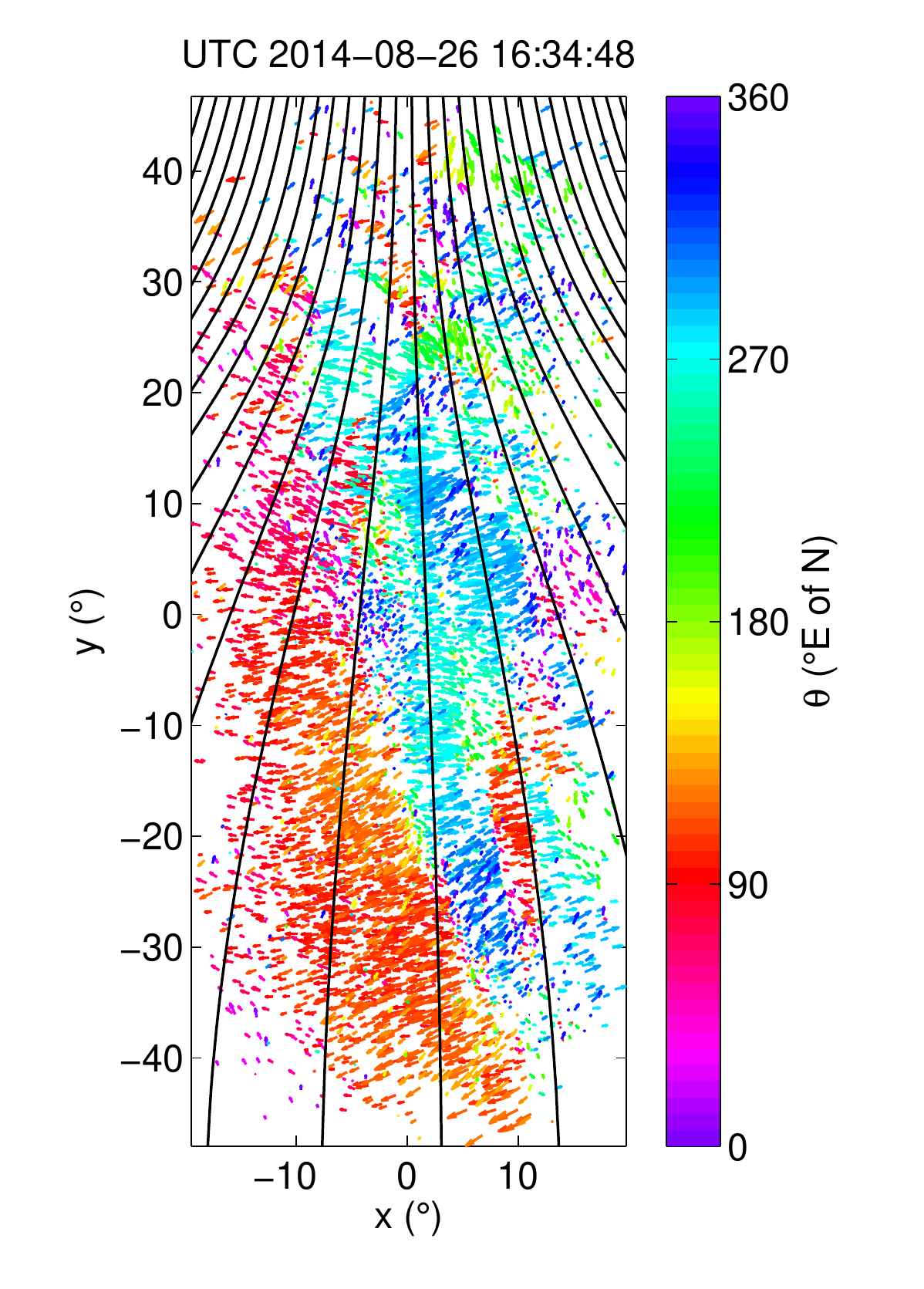}
  \includegraphics[width=0.49\textwidth]{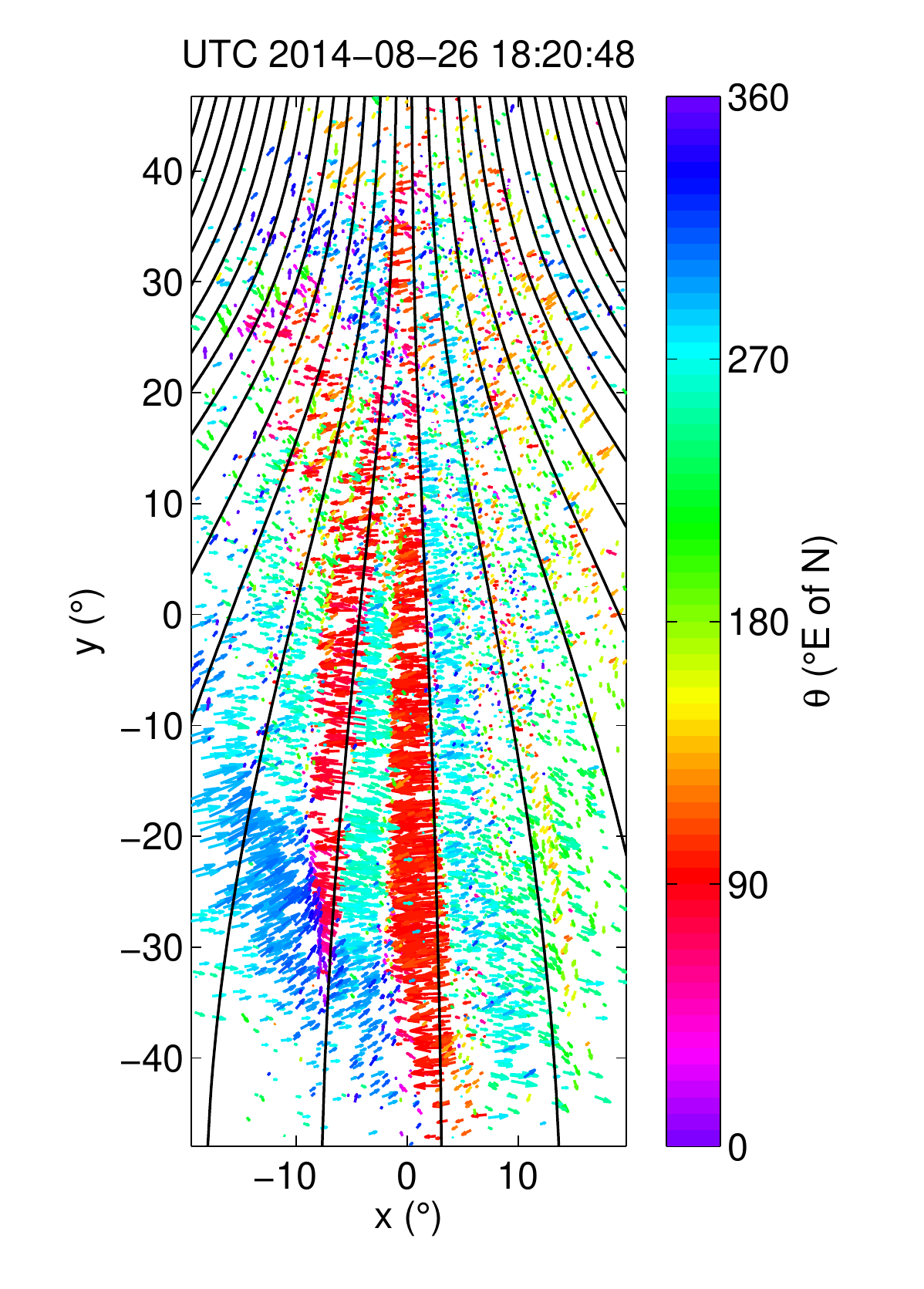}
  \caption{Snapshots of the angular distortion (equivalently $\nabla_\perp$TEC) vector field at two different times, constructed by combining three sequential pointings each integrated over 2\,min. Coordinates are angular distance across the MWA FoV, with north up and east to the left as per the astronomical convention (looking up at the sky). A cyclic colour scheme has been used to convey the position angle of each arrow, measured anticlockwise east of north as per the astronomical convention (so that e.g.~red and cyan corresponding to eastward and westward, respectively). Arrows are scaled to 100 times the actual angular displacement distance. Black solid lines are the geomagnetic field lines from $L = 1.8$, whose angular positions were computed assuming a simple dipole field model and a geomagnetic latitude of 38.6$^\circ$S for the MWA. Movie S1 contains an animation of the full dataset.}
  \label{fig:2014-08-26_arrow}
\end{figure*}

\subsection{Measuring Altitude}
Besides the detailed two-dimensional characterisation of the density distribution in the horizontal direction, a crude radial localisation of the irregularities is possible through parallax analysis. This involves forming a ``stereo'' set of images using two separate portions of the array, enabling a height triangulation based on the parallax shift of the distortion pattern. This was previously done by \citet{Loi2015_mn2e}, using an antenna-based division of the MWA. Here we perform a similar analysis, except that we divide the array not by antenna positions but by baseline midpoints. Since the ionospheric phase is measured on baselines rather than antennas, of which there are $O(N)$ times less than the number of baselines, many more visibilities can be included with a baseline-based split that has the same effective parallax baseline as an antenna-based split. This enhances the image sensitivity, increasing the number of detectable sources, and also improves positional accuracy, thereby improving the overall quality of the parallax measurement.

We used a simple ``greedy'' algorithm to construct the split. Although not necessarily optimal, the output is expected to be reasonable given the pseudo-random array layout of the MWA. Baselines were first sorted by east-west coordinate, then assigned to two groups in an alternating manner from the extreme east and west ends of the list. This was terminated when the centroid separation of the groups reached a threshold of 880\,m (the value associated with the split of \citet{Loi2015_mn2e}). A total of 2432 baselines were selected, around double the number (1261) used by \citet{Loi2015_mn2e} and about a third of the total number of baselines (8128). Figure \ref{fig:mwalayout} shows the MWA tile layout and the grouping of baselines used for stereo imaging.

Extraction of the parallax value proceeded identically to the method described in the appendix to \citet{Loi2015_mn2e}. A spatial interpolant over the east-west components of the refractive shifts measured for the full-array images was used as a reference model. For each image in the stereo set, we solved for the east-west displacement of the interpolant that minimised the sum of squared differences (weighted by source S/N ratio) between measured and model values. This was done with a grid search from $-30$ to $+30$\,arcmin in steps of 0.2\,arcmin. The parallax shift $p$ of the pattern was taken to be the angular separation of the best-fit displacements of the interpolant between a stereo pair of images. In the small-angle approximation the altitude is then $b/p$, where $b$ = 880\,m is the effective parallax baseline. 

\begin{figure*}[H]
  \centering
  \includegraphics[width=0.45\textwidth]{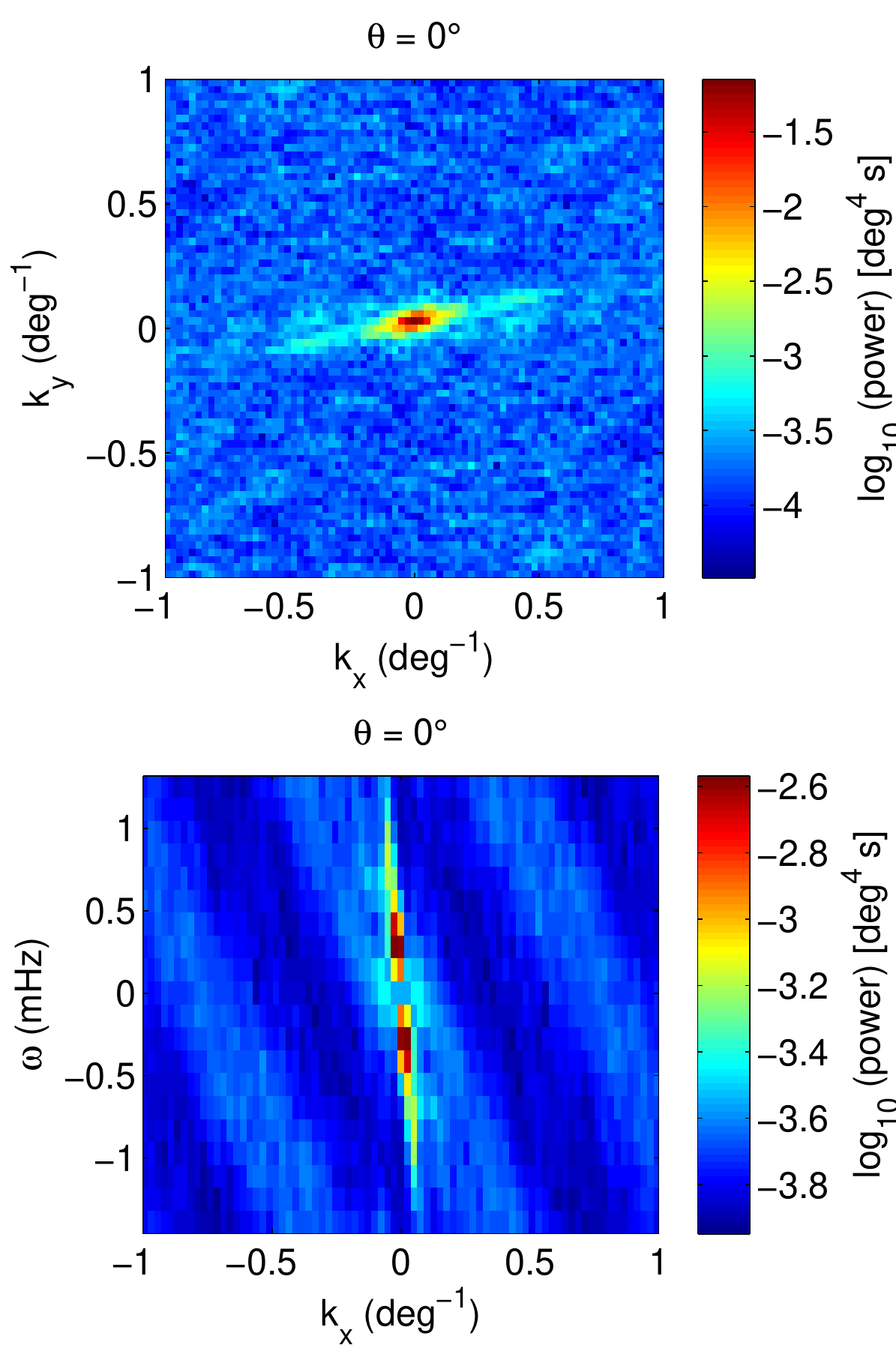}
  \includegraphics[width=0.45\textwidth]{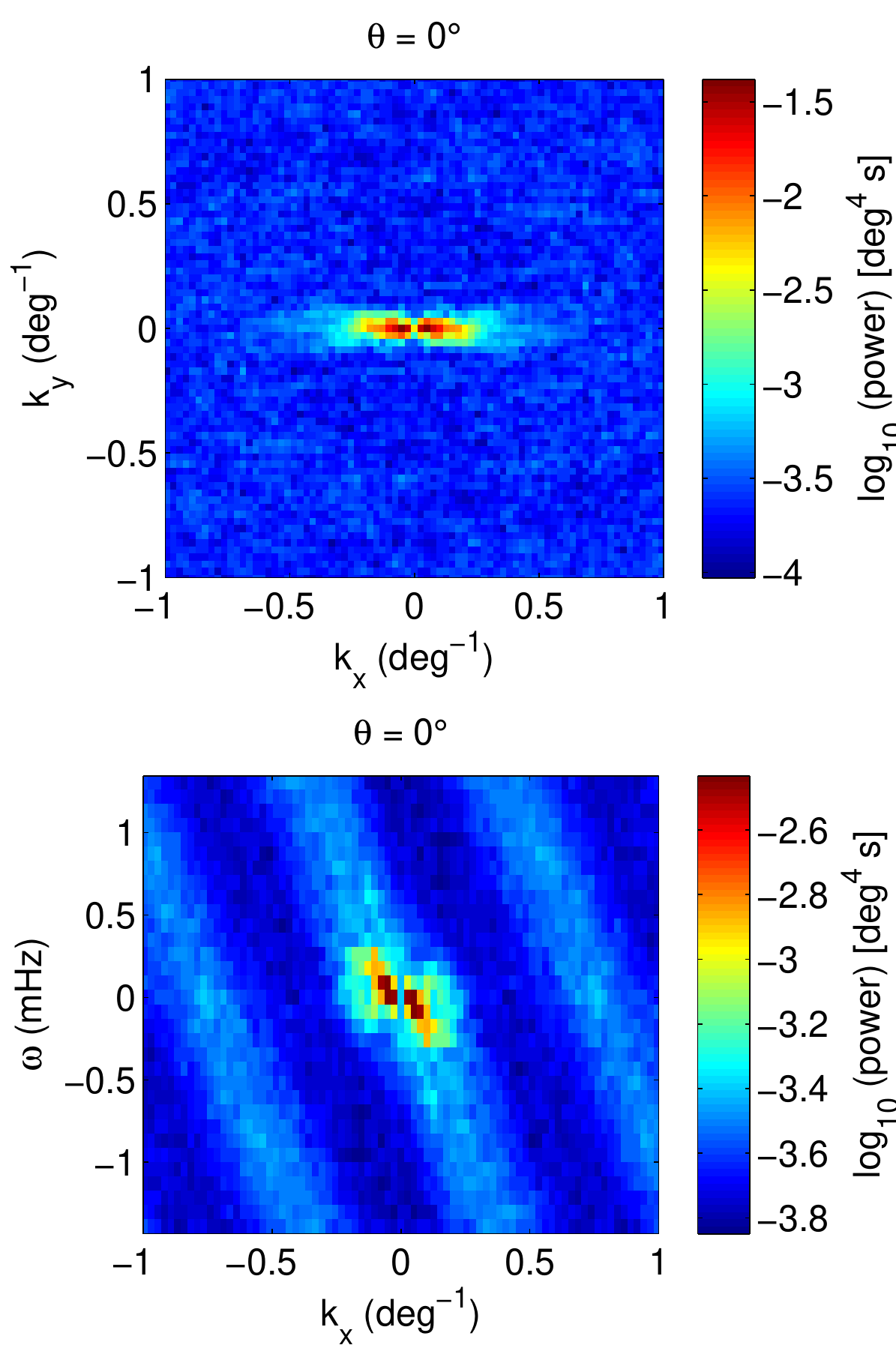}
  \caption{Spatiotemporal power spectrum of $\nabla_\perp$TEC measurements projected onto the E-W direction, for 1535--1735 UT (TID passage; left column) and 1735--2053 UT (duct formation; right column) on the 26th of August 2014. The spatial frequencies $k_x$ and $k_y$ are Fourier conjugates to $x$ and $y$ in Figure \ref{fig:2014-08-26_arrow}. The three-dimensional power spectral cube has been visualised by averaging along the $\omega$ (temporal frequency) axis in the top row, and along the $k_y$ axis in the bottom row.}
  \label{fig:2014-08-26_powerspec}
\end{figure*}

\section{Results and Discussion}\label{sec:results}
Animations of the distortion vector field for each of the three nights are shown in Movies S1, S2 and S3 (see \citet{Loi2015a_mn2e} for an explanation of the visualisation scheme). In all plots, black solid lines are the geomagnetic field lines from $L = 1.8$, computed assuming a simple dipole model with the MWA at a geomagnetic latitude of 38.6$^\circ$S and a magnetic declination of $0^\circ$ (this is more accurately $-0.153^\circ$ \citep{Thebault2015}, but we neglect this for simplicity of plotting).

\subsection{TID Event and Duct Formation}\label{sec:results_26th}
The data collected on the night of the 26th span 1100--2050 UT or 1900--0450 AWST (Australian Western Standard Time), beginning an hour after local sunset and ending an hour before local sunrise. From 1900--2330 AWST the TEC distribution is smooth and featureless, but a large-amplitude TID appears overhead between 2330--0100 AWST travelling northwest (snapshot in the left panel of Figure \ref{fig:2014-08-26_arrow}). Besides the large-scale undulations are finer-scale structures that materialise ahead of the main front as it sweeps overhead. Because of the cyclic pointing strategy, different parts of the sky are updated at different times. The time stamp shown in the header of Figure \ref{fig:2014-08-26_arrow} corresponds to the zenith pointing (middle one-third), which leads the southern pointing (bottom one-third) by 2\,min and the northern pointing (top one-third) by another 2\,min. This staggered update sequence can be seen in Movie S1, which contains the remaining data for the 26th.

Following the passage of the TID, a collection of regional-scale, field-aligned, duct-like density structures rapidly forms, becoming reasonably well-defined around 0.5\,hr after the TID passed and reaching their peak prominence near 0200--0300 AWST. A snapshot of these is shown in Figure \ref{fig:2014-08-26_arrow} (right), where the time stamp in the header is associated with the bottom one-third (which leads the middle one-third by 2\,min and the top one-third by another 2\,min). They persist until the end of the observations on the 26th and exhibit complex differential drift motions. Measurements of quiet-time, mid-latitude thermospheric wind speeds indicate values of 10--100\,m\,s$^{-1}$ \citep{Lloyd1972}, corresponding to angular drift speeds of the order 0.1--1$^\circ$\,min$^{-1}$ for ground-based observers. This is similar to the angular speed of the drifts seen in Movie S1, and suggests as one explanation that the observed motions may be driven by wind currents in the more collisional lower ionosphere, down to where the feet of the ducts possibly extend.

We measured the altitude of the TID to be $350 \pm 40$\,km by averaging the best-fit parallax over 2350--0120 AWST, where the uncertainty reflects the standard error of the parallax values computed over 28 independent fits (14 epochs $\times$ 2 instrument polarisations). Ionosonde data from the nearby observatory at Learmonth indicate that the altitude of the F region peak was around 230\,km, placing the TID in the topside region. The best-fit altitude of the ducts, averaged over 0120--0450 AWST for the zenith pointing, is very similar (also $350 \pm 40$\,km, from 68 independent fits). The proximity in time between the appearance of the TID and the ducts, as well as their spatial coincidence, suggest as one possible interpretation that the TID may have been a causal trigger for the ducts. This is reminiscent of the \citet{Cole1971} mechanism, where electron density irregularities (e.g.~from gravity waves) generate irregularities in the electric field that map along geomagnetic field lines, producing density ducts by flux-tube interchange. However the exact mechanism linking the TID with the ducts, and indeed whether the correlation seen here reflects causation, random coincidence or a common cause, cannot be ascertained solely based on the available data.

\begin{figure}[H]
  \centering
  \includegraphics[width=\columnwidth]{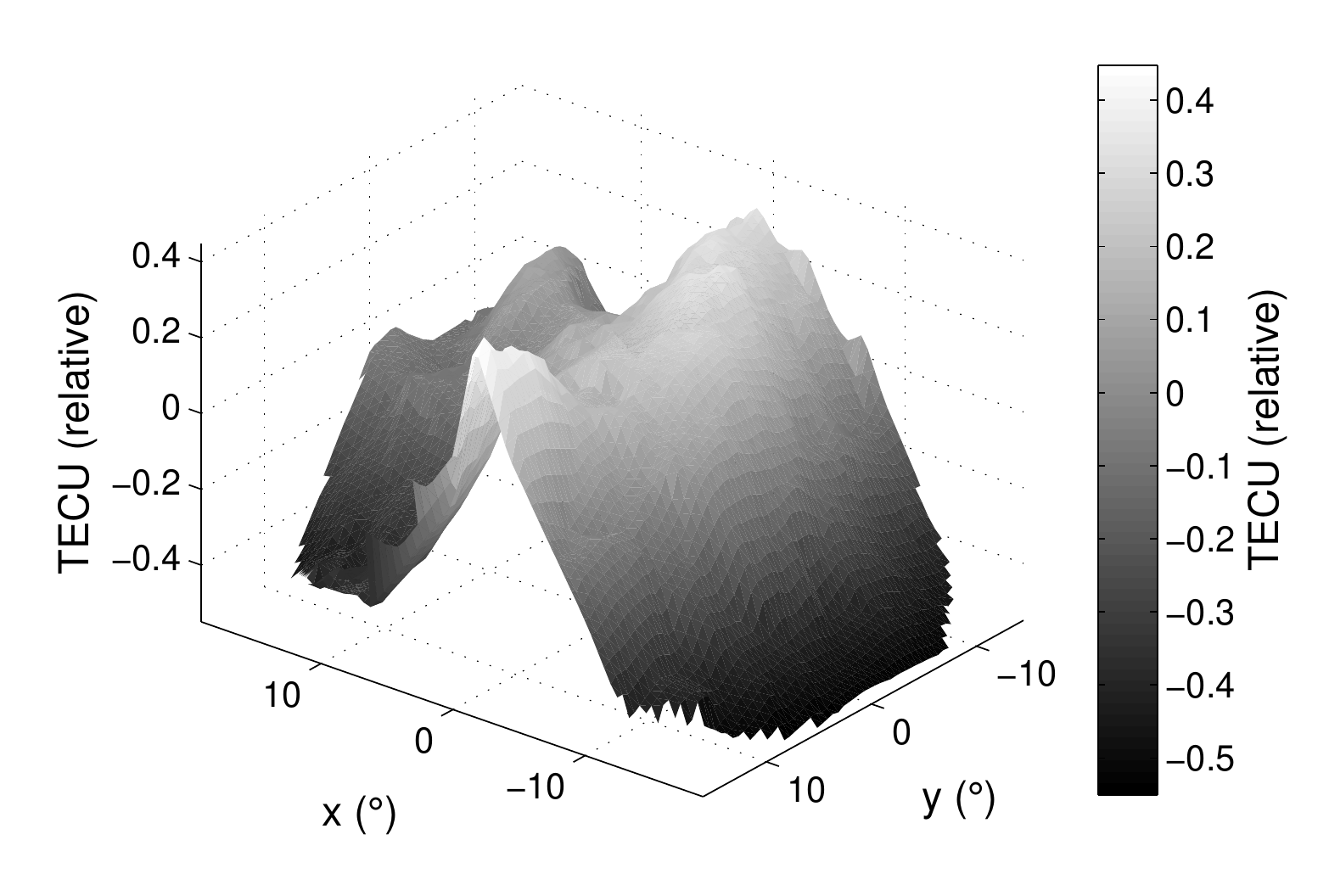}
  \includegraphics[width=\columnwidth]{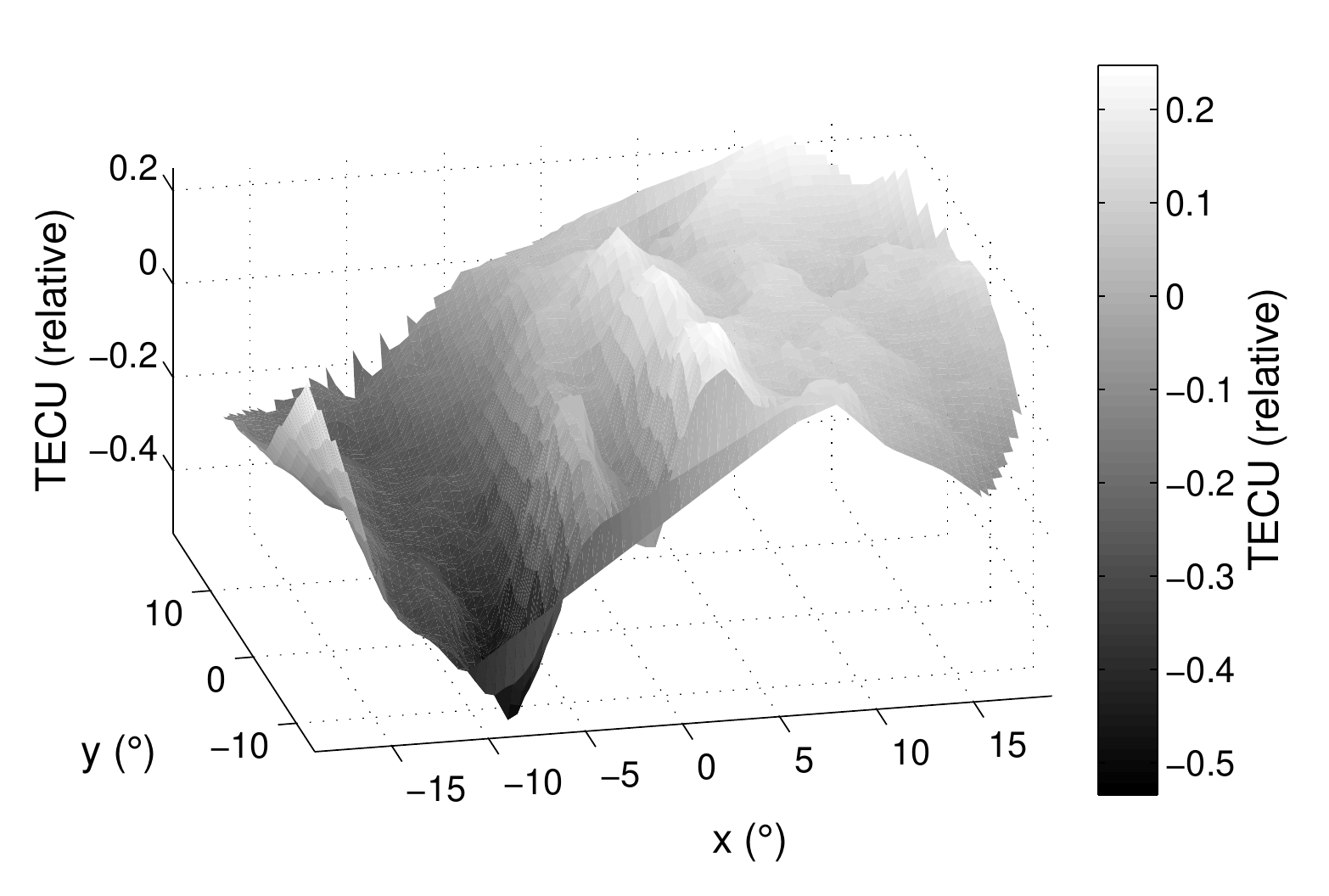}
  \caption{Surface plot of the relative TEC obtained by integrating the $\nabla_\perp$TEC vector fields for the zenith (middle) pointings at the two epochs shown in Figure \ref{fig:2014-08-26_arrow}. Note that the TEC can only be computed up to an arbitrary additive constant, since the quantity directly measured is a gradient. Associated errors are $\sim$0.05\,TECU. The integration algorithm to obtain TEC from $\nabla_\perp$TEC and the error analysis approach are described in Appendix \ref{sec:relTEC}.}
  \label{fig:2014-08-26_relTEC}
\end{figure}

The altitudes of structures aligned along the geomagnetic field should vary in accordance with the magnetic inclination ($-60^\circ$ at the MWA site). However, a separate analysis of the three pointings reveals no significant variation in the best-fit altitude values. This may be explained by the presence of significant structure in the radial direction, e.g.~the existence of ducts with feet at a range of latitudes. In this situation, the average altitude of irregularities along the line of sight would be roughly independent of direction. The complex ensemble of features in the data that appear to criss-cross and drift through one another supports this idea. Given that the extent of the TID exceeds the MWA FoV, it is plausible that if duct formation were to have been triggered by the TID, then the feet of the ducts should extend over a similarly large region, consistent with this picture.

Power spectrum analysis of the $\nabla_\perp$TEC vector field allows velocities and azimuths to be measured, as detailed in \citet{Loi2015a_mn2e}. The power spectra for two time intervals, one corresponding to the passage of the TID and the other to the period of duct formation, are shown in Figure \ref{fig:2014-08-26_powerspec}. These were computed for the east-west component of the $\nabla_\perp$TEC field. The propagation azimuths of the disturbances can be determined from the top row of panels ($k_x$-$k_y$ plane), and east-west phase speeds from the bottom row of panels ($\omega$-$k_x$ plane), noting that $x$ points west and $y$ points north. The diagonal ringing seen in the $\omega$-$k_x$ panels is a response feature induced by the rotation of the Earth (see Appendix \ref{sec:response}). The TID propagates towards N80$^\circ$W (280$^\circ$T) at a speed of $\sim$100\,m\,s$^{-1}$ and its wavelength is of order the east-west extent of the MWA FoV, which is $\sim$200\,km at 350\,km altitude. These properties are typical of medium-scale TIDs, which are driven by gravity waves, and indeed their spatiotemporal signatures are often seen in MWA data. 

\begin{figure*}[H]
  \centering
  \includegraphics[clip=true, trim=1cm 0cm 1cm 1.6cm, width=0.85\textwidth]{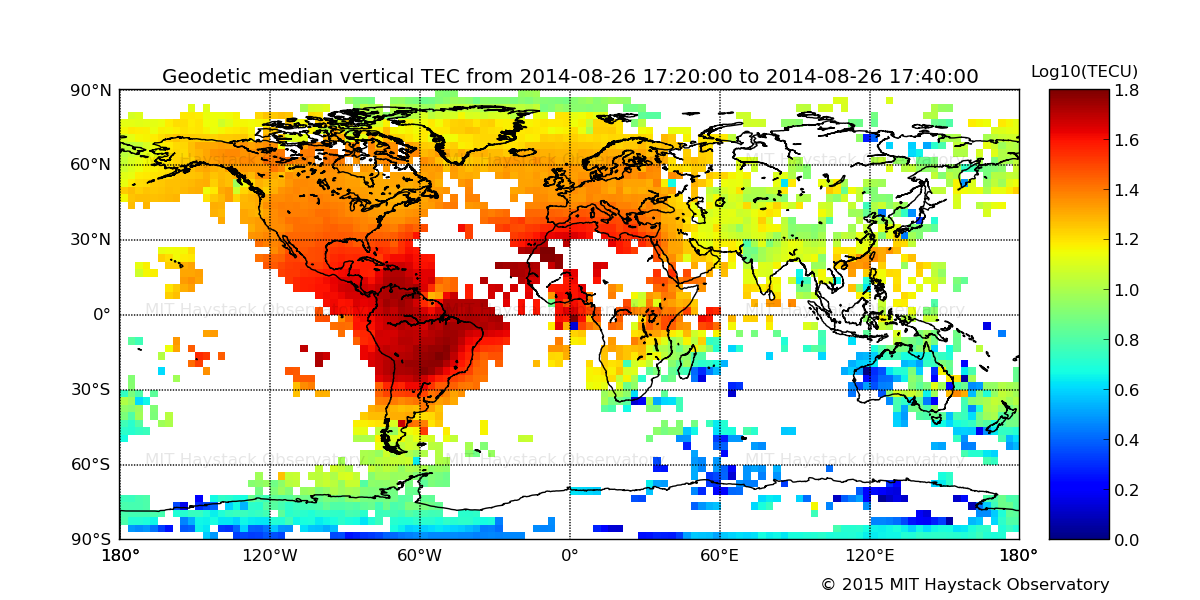}
  \caption{MIT Haystack MAPGPS analysis \citep{Rideout2006} of global distribution of TEC near local midnight on the 26th of August 2014, just prior to the passage of the TID over the observatory. Background TEC values are $\sim 2 \pm 2$\,TECU during this time near the MWATS survey area. The native MAPGPS global map product has a $1^\circ \times 1^\circ$ resolution in geodetic latitude and longitude, comparable to the physical size of the MWA FoV (cf.~Figure \ref{fig:geography}) at ionospheric heights. However, the resolution of the map plotted here is further averaged to around $3^\circ \times 3^\circ$ to emphasise large scale features.}
  \label{fig:global_tec}
\end{figure*}

The unusual aspect of this TID, however, is its amplitude. The steepest gradients measured in MWA data are around 10\,mTECU\,km$^{-1}$ (angular shifts of 30--40\,arcsec at 154\,MHz), but the TID seen here had 3--4 times steeper gradients, producing angular fluctuations of $\sim$2\,arcmin. Its peak-to-peak amplitude, measured by integrating the $\nabla_\perp$TEC field to obtain the relative TEC distribution (top panel of Figure \ref{fig:2014-08-26_relTEC}), is about 0.8\,TECU. GPS satellite maps of the global TEC distribution (Figure \ref{fig:global_tec}) indicate that the background level at the time was only $\sim 2 \pm 2$\,TECU, making this a significant fractional ($>$20\%) disturbance.

The cause of the TID is unknown. The spike in energetic proton flux on the 26th leads us to suggest that a proton precipitation event inducing substantial localised ionisation may have been responsible. Given that geomagnetic activity was very low during and prior to the appearance of the TID, alternate explanations may involve a lower-altitude energy source such as a meteor explosion occurring over central Australia. Unfortunately, conditions were cloudy for the cameras of the Desert Fireball Network (P.~Bland 2015, personal communication), and the low population density within most of the Australian continent means that records of such an event are likely non-existent.

\begin{figure}[H]
  \centering
  \includegraphics[clip=true, trim=0cm 0.5cm 0cm 0cm, width=\columnwidth]{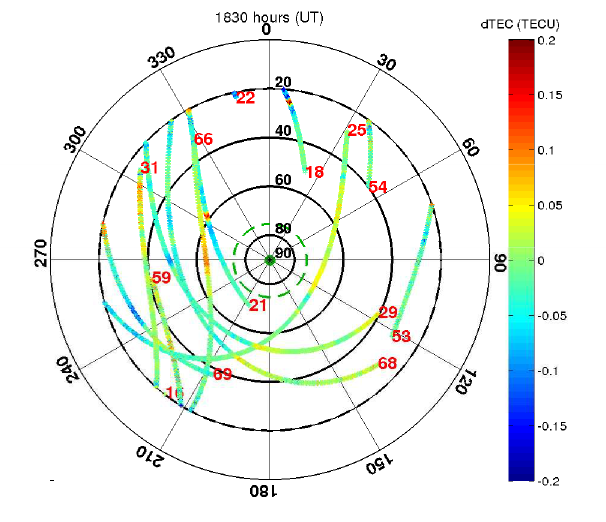}
  \caption{An MWA-centric azimuth-elevation plot of differential STEC values measured along individual line-of-sight tracks of GPS and GLONASS satellites visible at 1830 UT on the 26th of August 2014, at a cadence of 30\,s. The measurement precision is $\sim$0.01\,TECU. The dashed green circle of diameter 30$^\circ$ represents the MWA FoV, which is roughly circular with a FWHM extent of $\sim 30^\circ$ at 154\,MHz. Small-scale structure is evident in the form of coloured bumps along the tracks of individual satellites. Field-aligned ducts are prominent within the MWA FoV at this time (cf.~Figure \ref{fig:2014-08-26_arrow}, right panel). Movie S4 contains the data for the rest of the night.}
  \label{fig:gps_los}
\end{figure}

The power spectra of the ducts (right panels of Figure \ref{fig:2014-08-26_powerspec}) indicate dominantly east-west fluctuations. Their average drift speed, measured from the bottom-right panel of Figure \ref{fig:2014-08-26_powerspec}, is of the order several m\,s$^{-1}$ towards the west. This slow speed suggests that they are in-situ stationary features. Together with the diffuseness of the power spectral density in the $\omega$-$k_x$ plane, indicating the lack of a well-defined phase speed, this is consistent with the complex drift patterns observed. Unlike the duct-like structures reported by \citet{Loi2015_mn2e}, those seen here do not exhibit spatial periodicity on scales smaller than the FoV. This suggests that a different mechanism may be responsible for their formation or, alternatively, that another process limits the periodicity.

The physical width of the ducts, as estimated from the $\sim 10^\circ$ angular separation of the bands, is about 60\,km at a height of 350\,km. Electric field fluctuations on scales below about 10\,km are largely attenuated in the magnetosphere \citep{Reid1965}, and so the observed widths of the irregularities are near the minimum scale on which geomagnetic field lines act as equipotentials. It is therefore likely that the structures extend along the field lines into the other hemisphere, although observational limits here prevent direct confirmation of this. The relative TEC distribution (Figure \ref{fig:2014-08-26_relTEC}, bottom) indicates a TEC fluctuation of $\sim$0.4\,TECU associated with the ducts which, like for the TID, is a significant fraction of the background. Although the length scales and relative TEC variations are below the resolution of global TEC maps, cross-validation is possible with data from individual satellite tracks, which have higher precision and spatial resolution. Figure \ref{fig:gps_los} shows data from a Geoscience Australia GNSS receiver located at the observatory site, and Movie S4 shows data for the full night. Fine-scale features are apparent, and their angular scales (several degrees) and relative TEC variations (several tenths of a TECU) are in good agreement with MWA observations. However, the coverage is too sparse to provide the detailed characterisation possible with the MWA.

\begin{figure*}[H]
  \centering
  \includegraphics[clip=true, trim=0cm 0cm 0cm 0.5cm, width=0.9\textwidth]{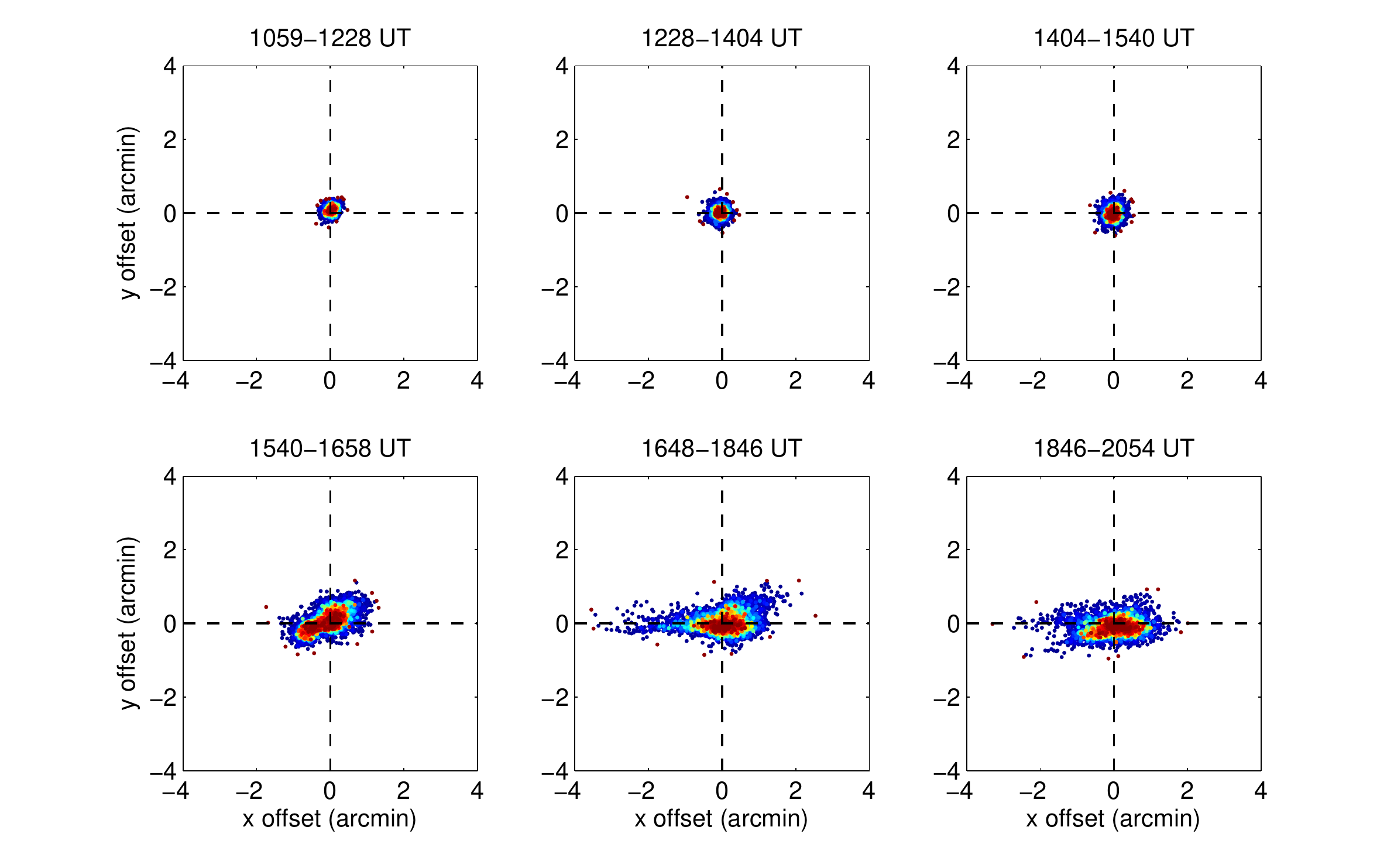}
  \caption{Scatter plots of position offsets measured in the southern pointing of the dataset taken on the 26th of August, divided into six sub-intervals of time. Points are coloured by the local density on the page to show the shape of the distribution, and only one in 5 points is plotted. The TID appears near 1600 UT (bottom-left panel), and the ducts reach their peak prominence at around 1800 UT (middle panel of bottom row), persisting till the end of the observation. The $x$ and $y$ directions are as for Figure \ref{fig:2014-08-26_arrow} (north up, east left). Note that angular offsets are in the direction of decreasing TEC, implying that the steepest TEC gradients associated with duct formation occur on the eastern walls of density enhancements (western walls of density depletions).}
  \label{fig:posoffsets}
\end{figure*}

A physical property that has not been reported so far in other MWA data is the pronounced asymmetry in the density gradients between the eastern and western walls of the ducts. Scatter plots of the angular offsets measured for the southern MWATS pointing are shown in Figure \ref{fig:posoffsets}, where the two rightmost panels on the bottom row correspond to when the ducts appear. It can be seen that the sizes of eastward offsets significantly exceed those to the west. This implies the steepest westward density gradients, i.e.~the eastern walls of density enhancements (western walls of density depletions) are sharper and more well defined. Asymmetries of this description have been reported in airglow observations of equatorial plasma bubbles \citep{Weber1978}, where the leading edges of density depletions were observed to be sharper than the trailing edges. However it is uncertain whether the same process is responsible for these asymmetries.

\subsection{Ducts and geomagnetic activity}

\subsubsection{27 August 2014 (Geomagnetic Storm)}\label{sec:results_27th}
\begin{figure*}[H]
  \centering
  \includegraphics[width=0.45\textwidth]{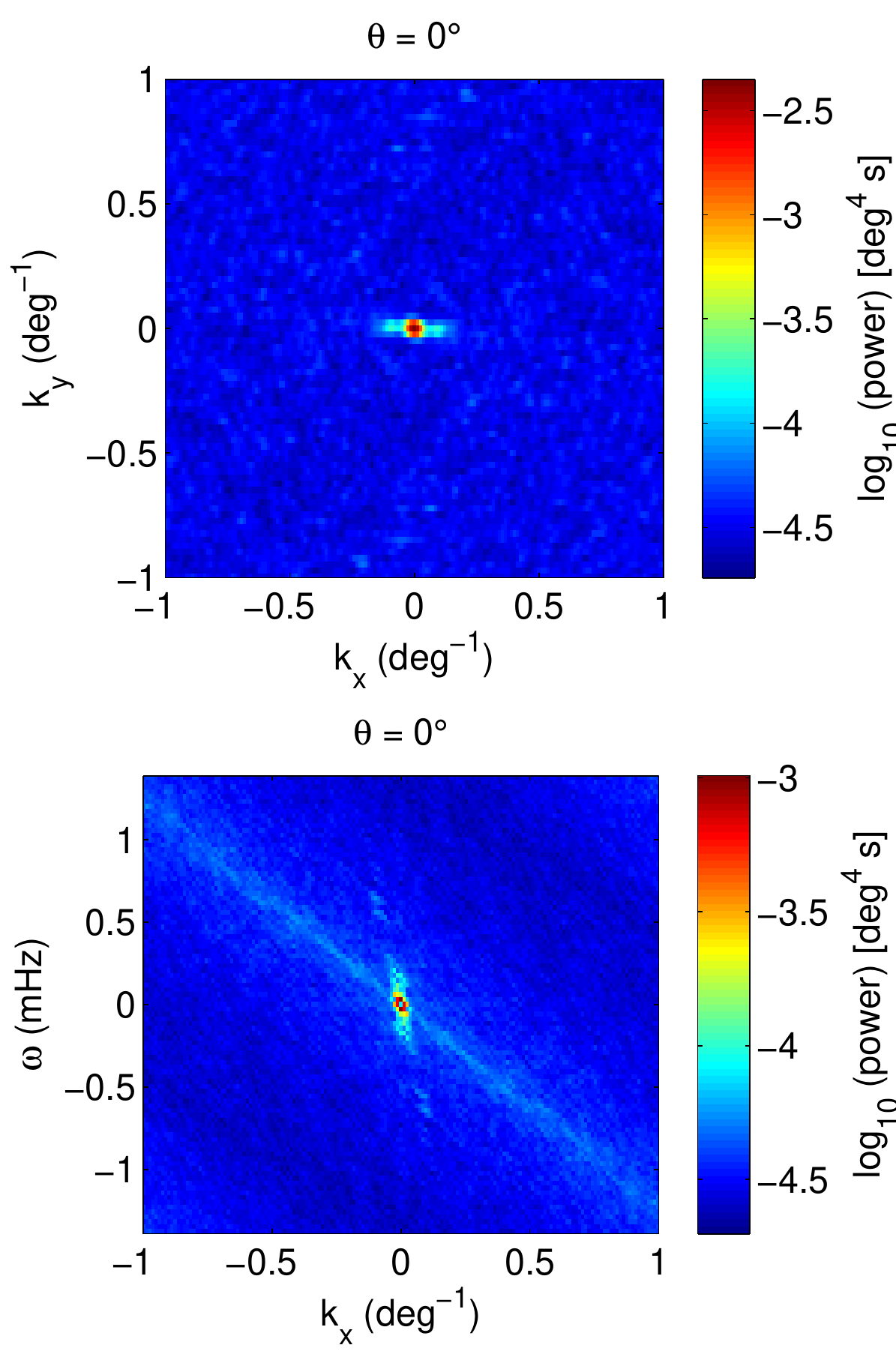}
  \includegraphics[width=0.5\textwidth]{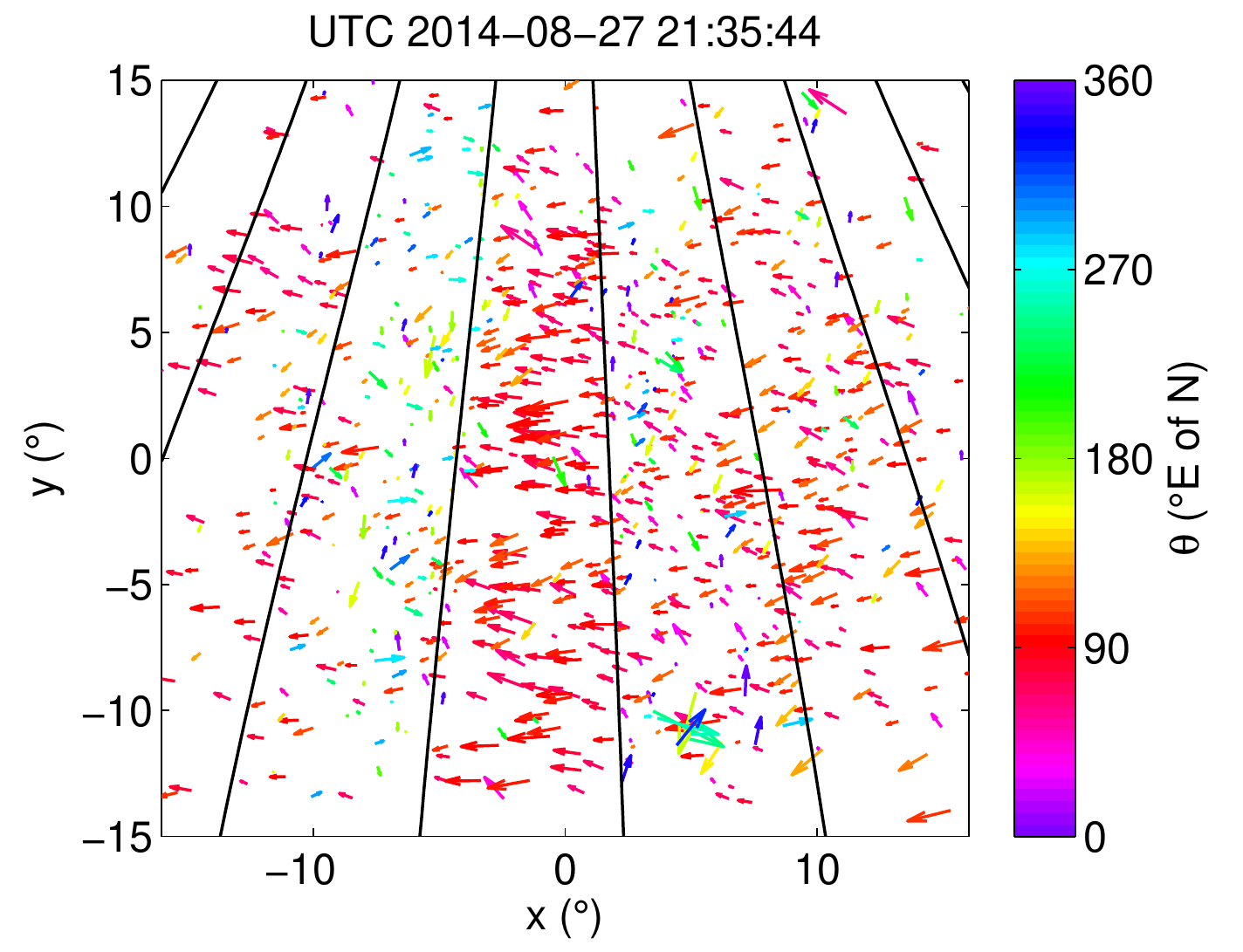}
  \caption{Left: Power spectrum for the interval 1851--2208 UT on the 27th of August 2014. Right: Distortion vector field for a single snapshot during this interval, showing regional-scale organisation into field-aligned structures. Arrows are scaled to 200 times the actual displacement distance. Movie S2 contains an animation of the full dataset.}
  \label{fig:2014-08-27_results}
\end{figure*}

The next interval spans 1450--2210 UT (2250--0610 AWST), around half a day after the onset of a geomagnetic storm. Movie S2 shows the time-lapse of the position offset vector field. The complicated structure observed on the previous night is no longer present, but whether this is reflects storm-induced ionospheric restructuring or a diffusive timescale shorter than $\sim$10\,hr is unknown. The diminishment in gradients associated with the ducts observed on the previous night (compare the two last panels of Figure \ref{fig:posoffsets}) supports the latter possibility. The TEC distribution on length scales smaller than the FoV is mostly structureless, but regional-scale field-aligned irregularities appear near the end of the observations (Figure \ref{fig:2014-08-27_results}, right panel) around 2100--2200 UT (0500--0600 AWST). That these occur in the second sweep of the telescope but not the first constrains the timescale of their growth to be under 3--4\,hr. 

The cluster of long, randomly-oriented arrows near the south-west of the FoV in Figure \ref{fig:2014-08-27_results} during the second pass corresponds not to a small-scale density irregularity drifting with the FoV but the celestial source Fornax A, a bright radio galaxy whose complicated structure is resolved in MWA observations. Difficulties with deconvolving the instrument response for this source have led to the appearance of spurious artefacts that have been erroneously cross-matched with real sources. However, a telltale indication that this not a terrestrial phenomenon is that this feature is fixed with respect to the celestial sphere.

The $\nabla_\perp$TEC power spectrum computed for the second half of the interval is shown in the left panels of Figure \ref{fig:2014-08-27_results}. The east-west orientation of the fluctuations is apparent in the top-left panel of Figure \ref{fig:2014-08-27_results} (the $k_x$-$k_y$ panel), and their characteristic drift speed is $\sim$40\,m\,s$^{-1}$ to the west.

\subsubsection{28 August 2014 (Storm Recovery)}\label{sec:results_28th}
Although the full time span of MWA data collected on the third day covers the same AWST range as the 27th, only about 10\% of these data have been imaged and analysed. This corresponds to the time period between 1650--1830 UT (0050--0230 AWST). The time-lapse of the position offset vector field is shown in Movie S3. Substorm activity is pronounced during this time, and a large negative spike in AL is noted to occur near 1700 UT. Unlike the previous night, fine-scale structuring is evident throughout the interval in the form of weak, narrow, field-aligned structures. A fast-moving TID also appears in the second half of the interval, propagating towards the NW. 

\begin{figure*}[H]
  \centering
  \includegraphics[width=0.45\textwidth]{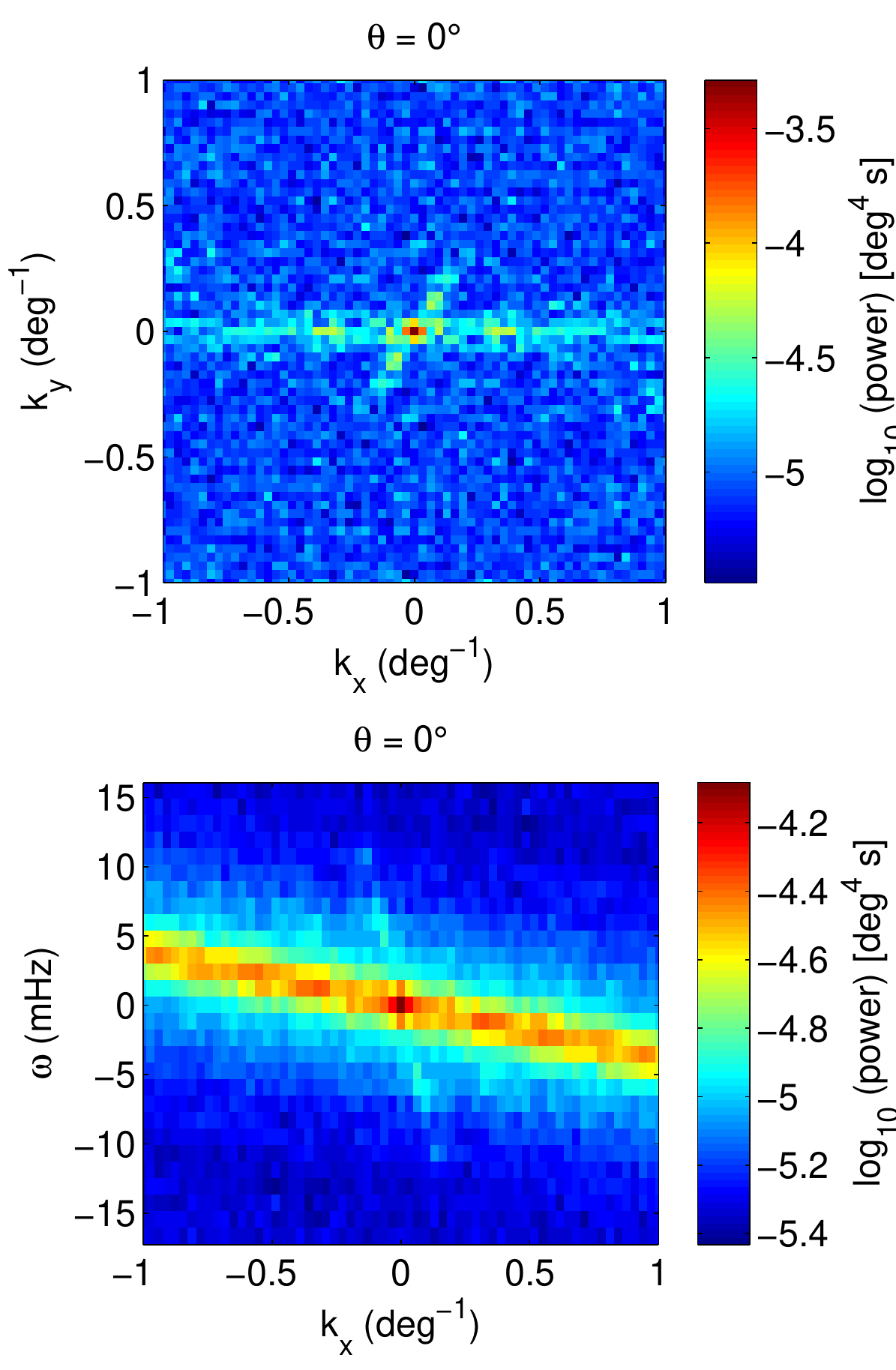}
  \includegraphics[width=0.5\textwidth]{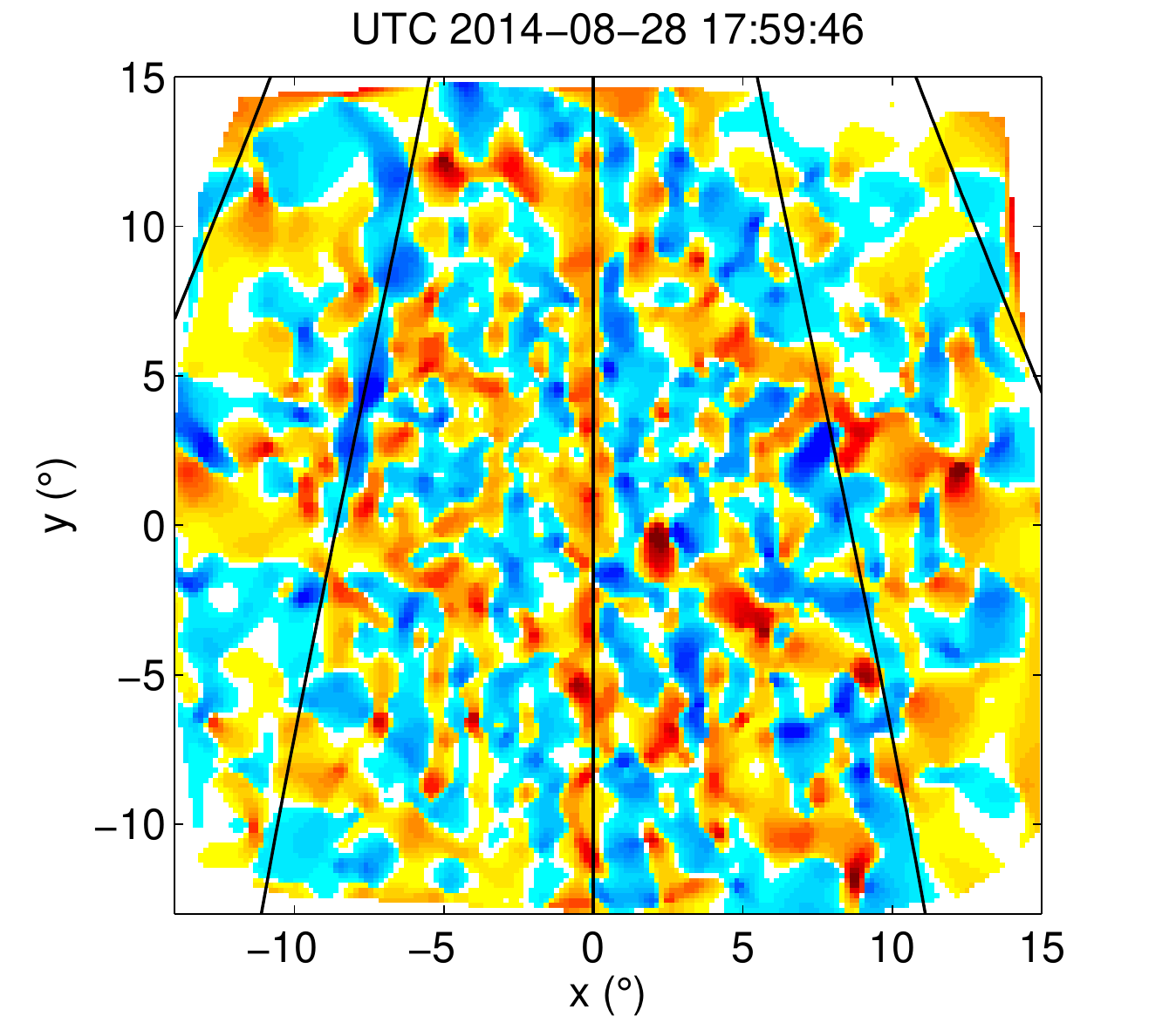}
  \caption{Left: Power spectrum for the interval 1740--1800 UT on the 28th of August 2014. Right: Divergence of the vector field for a single snapshot during this interval, showing weak field-aligned bands and also NW-propagating disturbance fronts. (Structures having angular scales close to the spatial sampling density are more clearly seen using a divergence rather than an arrow plot.) Movie S3 contains an animation for the full dataset.}
  \label{fig:2014-08-28_results}
\end{figure*}

The left panels of Figure \ref{fig:2014-08-28_results} show the power spectrum computed for the period during which the TID appears (0140--0200 AWST), and the right panel shows the divergence of the vector field for a snapshot image taken during this period. Signatures both of the field-aligned bands and the NW-propagating TID are visible in the top-left panel as two streaks angled at $\sim$45$^\circ$ to one another. The field-aligned bands appearing on this night are very similar to those commonly appearing in other MWA data in terms of their angular widths, spatial periodicity, extensiveness in the longitudinal direction, and associated $\nabla_\perp$TEC fluctuations.

The speed of the TID, both estimated from Movie S3 and the power spectrum, is about 300\,m\,s$^{-1}$ (for an altitude of 350\,km). The wavelength is about 40\,km (classifying this as a small-scale TID, or SSTID), and its oscillation period is $\sim$2\,min. These properties are consistent with the acoustic branch rather than the gravity wave branch of the dispersion relation for atmospheric waves, since the lower cutoff is 10--12\,min for gravity wave periods \citep{Hines1960}. Little is understood about the acoustic waves that drive SSTIDs, because the small length scales (several tens of kilometres) and subtle fluctuation amplitudes (several per cent of the background TEC) make them difficult to detect. However, the spatial scales to which the MWA is sensitive make it suitable for studying SSTIDs. Suspected sources of acoustic waves in the atmosphere are extreme tropospheric weather \citep{Georges1973} and auroral precipitation \citep{Campbell1963}. The speed of the TID and its direction of arrival are consistent with a high-latitude origin of the disturbance, possibly due to a particle precipitation event around 1700 UT.

\section{Conclusion}\label{sec:conclusion}
We have presented a case study reporting the formation of long-lived, field-aligned density ducts shortly after the passage of a TID. We suggest on the basis of spatial and temporal proximity that they are causally related, possibly evidencing the viability of the mechanism proposed by \citet{Cole1971}. Our results do not argue against the viability of other mechanisms, which may operate when their respective conditions are right. Three consecutive nights of MWA data were analysed, spanning 26--28 August 2014. No similar event has been seen in $\sim$50 hours of previously analysed MWA data, suggesting that such occurrences are fairly rare over the observatory. The origin of the TID itself (which had an unusually large amplitude) is unknown, although an abrupt spike in energetic proton levels on the day of the 26th suggests possible excitation by a proton precipitation event. Pronounced asymmetries were noted in the TEC gradients between the eastern and western walls of the ducts, a property for which we do not yet have an explanation. None of the structures that formed on the 26th remained present when substantial geomagnetic activity set in on the 27th, upon which the TEC distribution became smooth and featureless. Smaller-scale, periodically-spaced, field-aligned bands with properties similar to those often present in other MWA data appeared during the recovery phase on the 28th. An acoustic disturbance (an SSTID) with a possible high-latitude origin was also detected, its time of appearance consistent with propagation from a high-latitude substorm event. 

It is instructive to note that structure on 10--100\,km scales (within the extent of the MWA FoV) in this study appeared preferentially under quiet geomagnetic conditions, while a structureless ionosphere was seen during the most disturbed period. However, given the small amount of data analysed in this study, this does not necessarily indicate an anticorrelation between ionospheric and geomagnetic activity but rather that it may be difficult to forecast ionospheric behaviour on these scales based on global geomagnetic activity. Clear night-to-night variations in the types of structures present, and direct observations of the growth of features on hour-long timescales, also imply that short-term ionospheric forecasting (e.g.~to improve the accuracy of interferometric calibration) might be a complicated task. The possible anticorrelation between geomagnetic activity and the appearance of regional-scale (10--100\,km) TEC structures may be explained by an intrinsic drop in their occurrence rate, perhaps due to an increase in the characteristic length scale of persistent density fluctuations, or physical conditions being otherwise unfavourable. Alternatively, it may be that under geomagnetically disturbed conditions, the formation of regional-scale structures occurs over a much wider range of altitudes and/or more frequently, causing them to blend into one another and leading to a smooth TEC distribution.

This study demonstrates that geospace physics research, particularly the study of ionospheric structure on regional (10--100\,km) scales, can be a fully commensal application of the MWA alongside astronomical research. The refractive analysis described here requires only the total-intensity images, which are the end products used in many astronomical studies. While previous work \citep[e.g.][]{Loi2015_mn2e} has shown the value of the MWA's widefield snapshot capabilities for ionospheric research, the current study emphasises in addition the fast steering capabilities afforded by electronic beamforming, something which has not been possible with traditional, mechanically-steered telescopes. The raster scanning mode used in MWATS has roughly tripled the effective FoV of the MWA, making this study the broadest angular observation of the ionosphere by a radio telescope to date. Finally, we stress that although the MWA is being used mostly for radio astronomy, harnessing its unique capabilities as a geospace probe will be a necessary step towards realising its full scientific potential.


%
%
%

\appendix

\section{Computing relative TEC}\label{sec:relTEC}
The $\nabla_\perp$TEC vector field, although continuous in reality, is in practice only measured in a set of discrete directions given by the locations of radio sources in an image. These are randomly distributed to first approximation, with somewhat denser coverage towards the centre of the primary beam where the instrument is most sensitive. Noise, both from Gaussian position fitting errors and from errors in the reference positions, is also present in the vector field. Integrating such a vector field is non-trivial, in part because the randomness of the spatial sampling introduces an arbitrariness in the path of such an integration, and more importantly because of the loss of conservativeness (potentiality) of the vector field due to the introduction of noise. Put another way, the underlying TEC distribution is a scalar field, meaning that the integral of its derivative between any two points should be path-independent (i.e.~self-consistent), but this property is destroyed by measurement noise.

The objective is to find some reasonable approximation to the underlying scalar field (the TEC distribution) given a noisy, non-uniformly sampled gradient vector field (the $\nabla_\perp$TEC distribution) at some instance of time. Our approach to this was to average the TEC distribution obtained over many trials, where each trial involved looping over a random permutation of measurement points and successively piecing together the relative TEC distribution constructed in patches based on local gradient extrapolations. In more detail, the steps of the algorithm are as follows:
\begin{enumerate}
  \item Construct the Voronoi diagram of the measurement points, each of which has a two-dimensional spatial coordinate $(p_x, p_y)$ and an associated gradient vector $(v_x, v_y)$.
  \item Compute a random permutation of the Voronoi cells (each of which contains one measurement point), and begin looping over them to assign relative TEC values. We will refer to cells with no assigned value as being ``unfilled'', and those which do as being ``filled''.
  \item For the current cell $i$:
    \begin{enumerate}
    \item If this is unfilled, use the associated $\nabla_\perp$TEC vector $(v_x^i, v_y^i)$ to calculate the relative TEC values between it and any unfilled Voronoi neighbours $j$ by linear extrapolation, i.e.~take the TEC difference between cells $i$ and $j$ to be $\Delta \mathrm{TEC}_{ij} = v_x^i (p_x^j - p_x^i) + v_y^i (p_y^j - p_y^i)$.
      \item If any newly-filled cells are neighbours of filled cells, adjust the reference TEC values of each group of cells (identified by a certain group ID) in a manner that minimises the weighted difference between the actual and extrapolated TEC differences between adjacent cells with different group IDs, and then merge the group IDs. For each such pair of cells, the cell containing the source with higher signal-to-noise ratio (SNR) is used as the reference point for extrapolation, and each pair is weighted by the factor SNR/$d$, where $d$ is the Euclidean distance between the cells.
    \end{enumerate}
  \item At the end of the above loop, all cells will be filled. Save a copy of the relative TEC distribution, and reset all Voronoi cells to their unfilled state.
  \item Repeat steps 2--4 for a desired number of trials.
  \item Adjust the relative TEC values of each trial so that their medians are matched, and then compute the mean and standard deviation on a cell-by-cell basis over all trials.
\end{enumerate}
We have validated the accuracy of this algorithm using a number of sinusoidal test functions and sampling distributions borrowed from actual MWA data, where the original function was recovered to within errors. The errors were computed on a cell-by-cell basis as the standard deviation divided by the square root of the number of trials. Though not necessarily the most optimal or efficient approach, the results appear to be reasonable for reasonable inputs and so we have not invested additional time to further improve this.

\section{Power spectrum response functions}\label{sec:response}
\begin{figure*}[H]
  \centering
  \includegraphics[width=0.45\textwidth]{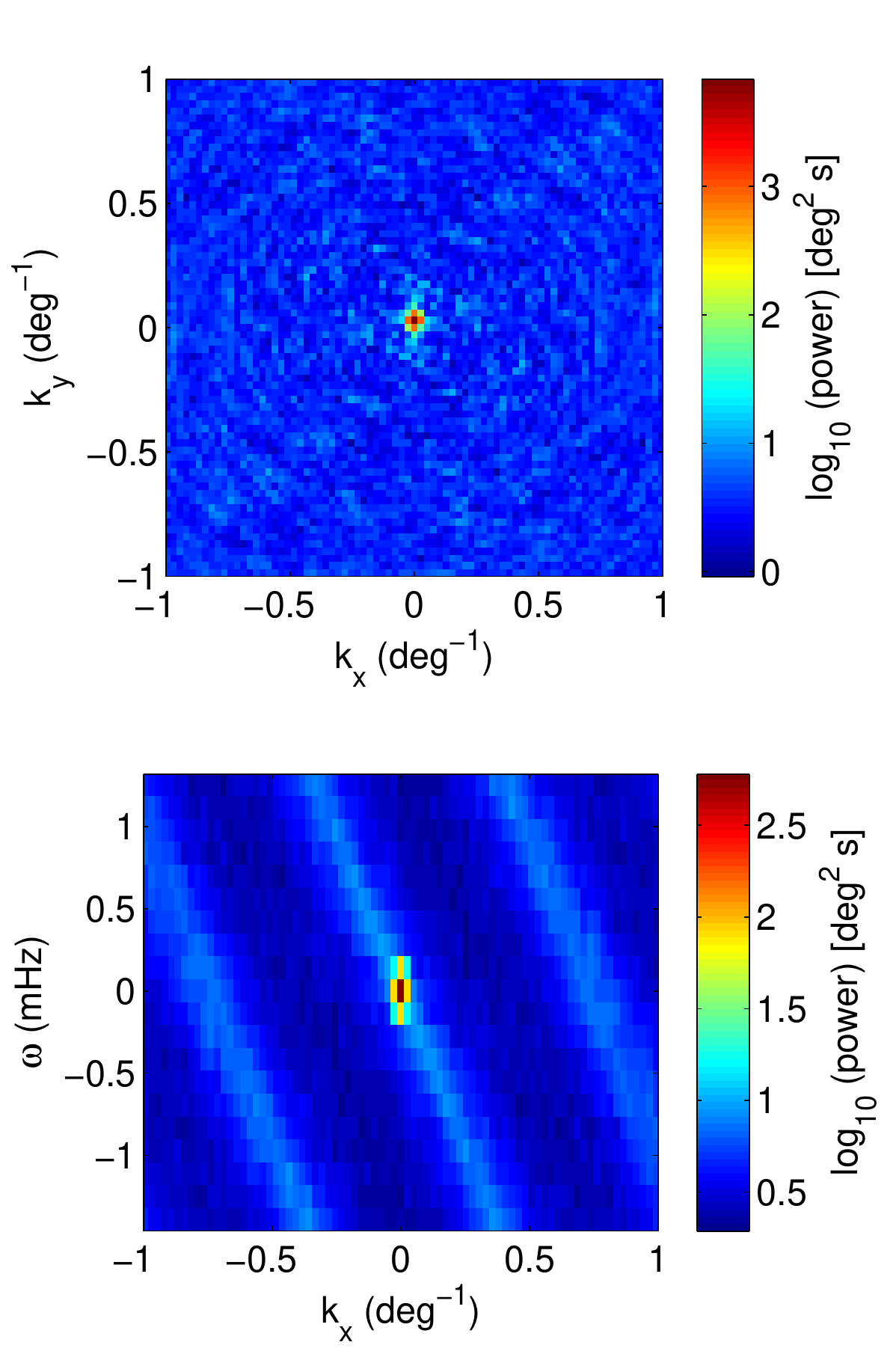}
  \includegraphics[width=0.45\textwidth]{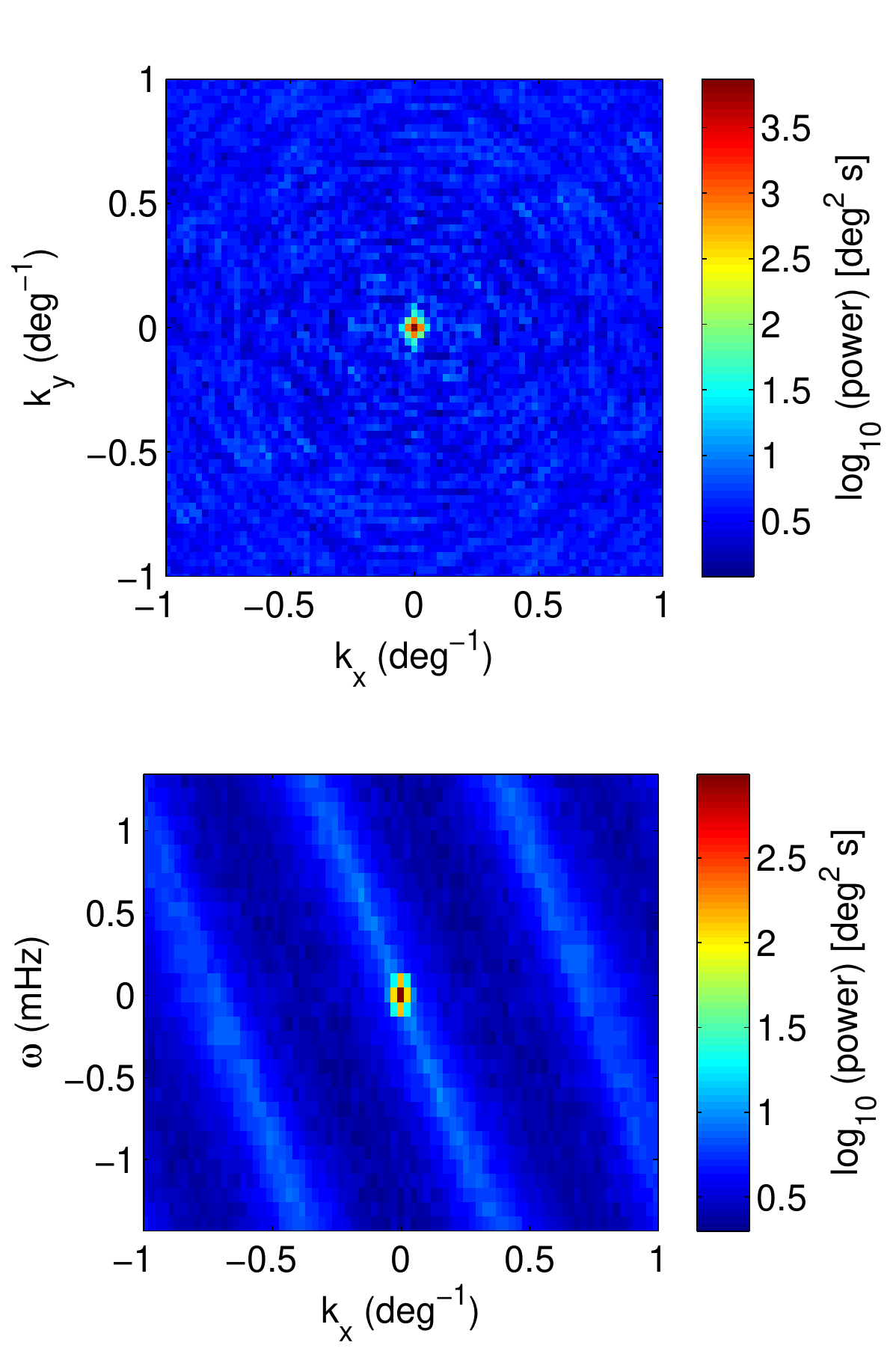}
  \caption{Response functions corresponding to the power spectra in Figure \ref{fig:2014-08-26_powerspec}.}
  \label{fig:MWATS_respfn}
\end{figure*}

\begin{figure*}[H]
  \centering
  \includegraphics[width=0.45\textwidth]{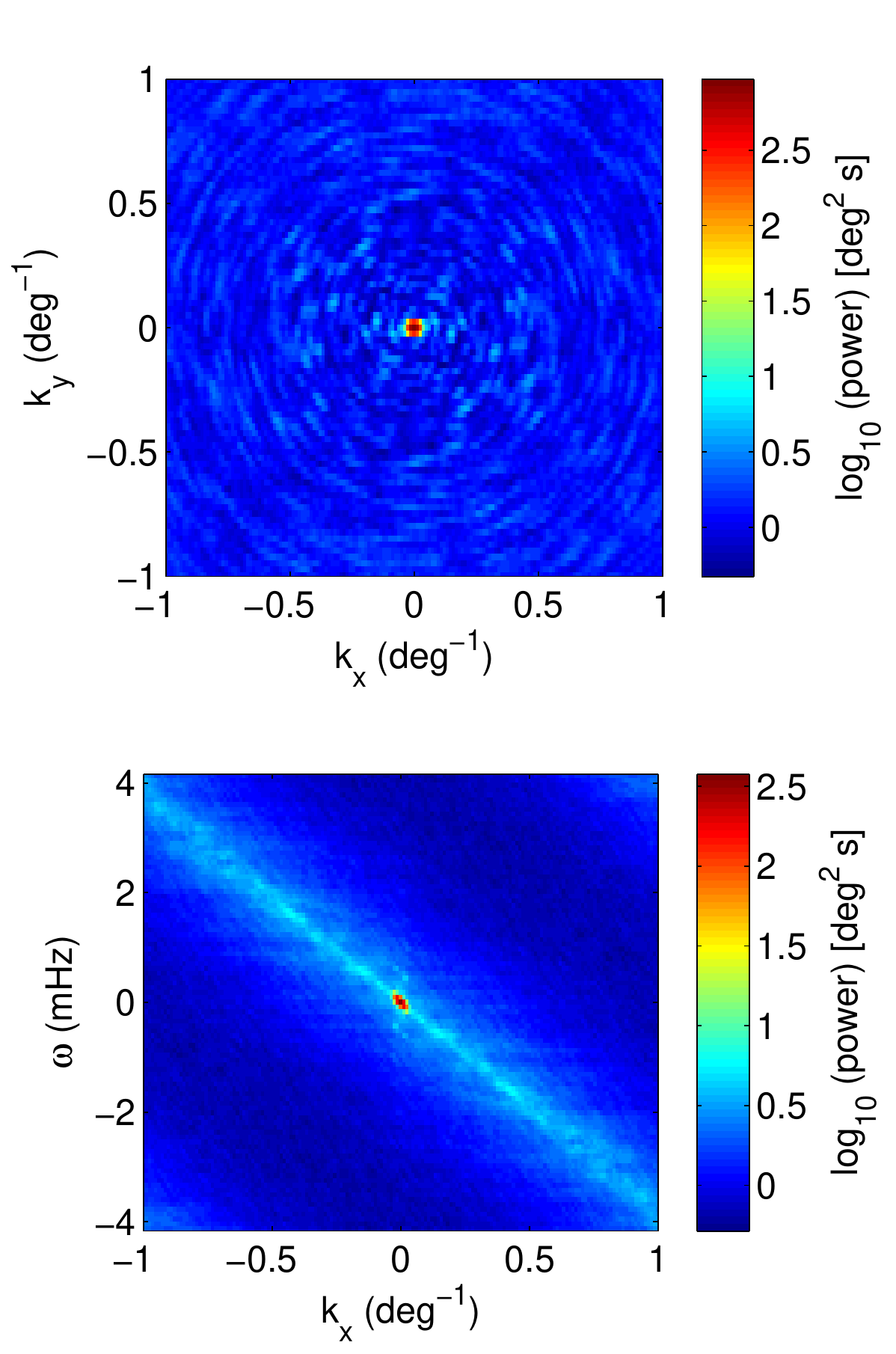}
  \includegraphics[width=0.45\textwidth]{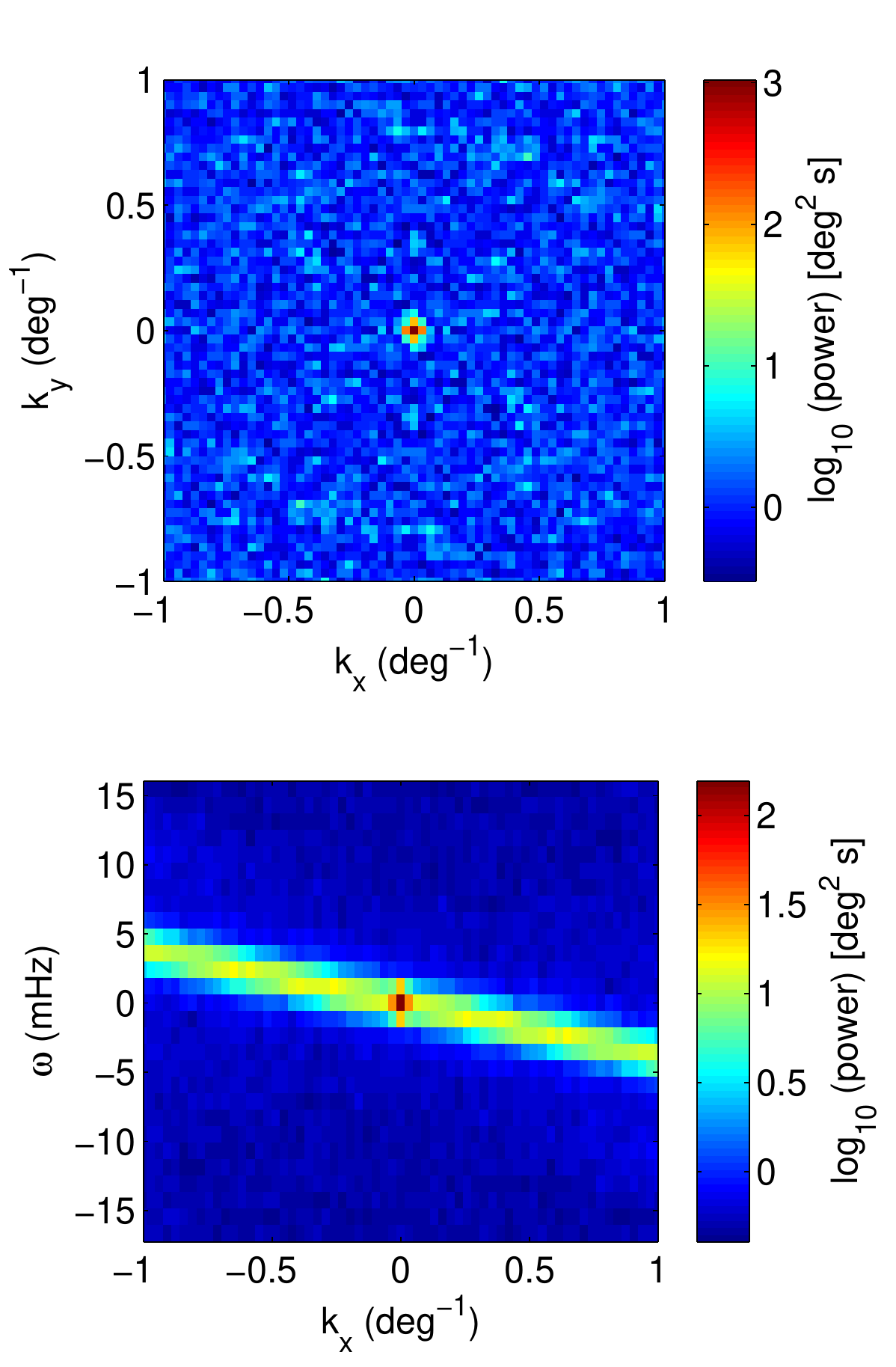}
  \caption{Response functions corresponding to the power spectra in the left panels of Figures \ref{fig:2014-08-27_results} and \ref{fig:2014-08-28_results}.}
  \label{fig:EoRTS_respfn}
\end{figure*}

The response functions for the power spectrum analysis are shown in Figures \ref{fig:MWATS_respfn} and \ref{fig:EoRTS_respfn}. These were obtained by taking the Fourier transform of the datacubes with all measurement values replaced by unity. Details of the method can be found in \citet{Loi2015a_mn2e}. The negative slope of the ringing features in the response functions arise from the drift of celestial sources as the Earth rotates, which produces diagonal tracks in the spatiotemporal sampling pattern. The slope of these features closely matches the angular rotation speed of the Earth at the geographical latitude of the MWA ($\sim 4 \times 10^{-3}$\,deg\,s$^{-1}$).

%
%
%
%

\begin{acknowledgments}
The OMNI data were obtained from the GSFC/SPDF OMNIWeb interface at http://omniweb.gsfc.nasa.gov. Global GPS TEC analysis at MIT Haystack Observatory is supported by US National Science Foundation Cooperative Agreement AGS-1242204 with the Massachusetts Institute of Technology. The MWA data supporting this paper are available on request submitted via email to the corresponding author at stl36@cam.ac.uk. This scientific work makes use of the Murchison Radio-astronomy Observatory, operated by CSIRO. We acknowledge the Wajarri Yamatji people as the traditional owners of the Observatory site.  Support for the operation of the MWA is provided by the Australian Government Department of Industry and Science and Department of Education (National Collaborative Research Infrastructure Strategy: NCRIS), under a contract to Curtin University administered by Astronomy Australia Limited. We acknowledge the iVEC Petabyte Data Store and the Initiative in Innovative Computing and the CUDA Center for Excellence sponsored by NVIDIA at Harvard University.
\end{acknowledgments}

\end{article}
%
%
%
%
%
%
%
%



\begin{thebibliography}{90}
\providecommand{\natexlab}[1]{#1}
\expandafter\ifx\csname urlstyle\endcsname\relax
  \providecommand{\doi}[1]{doi:\discretionary{}{}{}#1}\else
  \providecommand{\doi}{doi:\discretionary{}{}{}\begingroup
  \urlstyle{rm}\Url}\fi

\bibitem[{\textit{Abel and Thorne}(1998)}]{Abel1998}
Abel, B., and R.~M. Thorne (1998), {Electron scattering loss in the Earth's
  inner magnetosphere 1. Dominant physical processes}, \textit{JGR},
  \textit{103}(A2), 2385--2396.

\bibitem[{\textit{Angerami}(1970)}]{Angerami1970}
Angerami, J.~J. (1970), {Whistler duct properties deduced from VLF observations
  made with the Ogo 3 satellite near the magnetic equator}, \textit{JGRA},
  \textit{75}(31), 6115 -- 6135, \doi{10.1029/JA075i031p06115}.

\bibitem[{\textit{Bernhardt and Siefring}(2006)}]{Bernhardt2006}
Bernhardt, P.~A., and C.~L. Siefring (2006), {New satellite-based systems for
  ionospheric tomography and scintillation region imaging}, \textit{Radio
  Science}, \textit{41}(5), RS5S23, \doi{10.1029/2005RS003360}.

\bibitem[{\textit{Booker}(1979)}]{Booker1979}
Booker, H.~G. (1979), {The role of acoustic gravity waves in the generation of
  spread-F and ionospheric scintillation}, \textit{JATP}, \textit{41}, 501 --
  515, \doi{10.1016/0021-9169(79)90074-6}.

\bibitem[{\textit{Bougeret}(1981)}]{Bougeret1981}
Bougeret, J.~L. (1981), {Some Effects Produced by the Ionosphere on Radio
  Interferometry: Fluctuations in Apparent Source Position and Image
  Distortion}, \textit{A\&A}, \textit{96}, 259--266.

\bibitem[{\textit{Bowman et~al.}(2013)}]{Bowman2013_mn2e}
Bowman, J.~D., et~al. (2013), {Science with the Murchison Widefield Array},
  \textit{PASA}, \textit{30}(2013), e31, \doi{10.1017/pas.2013.009}.

\bibitem[{\textit{Burgess and Inan}(1993)}]{Burgess1993}
Burgess, W.~C., and U.~S. Inan (1993), The role of ducted whistlers in the
  precipitation loss and equilibrium flux of radiation belt electrons,
  \textit{JGR}, \textit{98}(A9), 15,643 -- 15,665, \doi{10.1029/93JA01202}.

\bibitem[{\textit{Calvert and Warnock}(1969)}]{Calvert1969}
Calvert, W., and J.~M. Warnock (1969), Ionospheric irregularities observed by
  topside sounders, \textit{Proceedings of the IEEE}, \textit{57}, 1019--1025,
  \doi{10.1109/PROC.1969.7146}.

\bibitem[{\textit{Campbell and Young}(1963)}]{Campbell1963}
Campbell, W.~H., and J.~M. Young (1963), {Auroral-zone observations of
  infrasonic pressure waves related to ionospheric disturbances and geomagnetic
  activity}, \textit{JGR}, \textit{68}(21), 5909 -- 5916,
  \doi{10.1029/JZ068i021p05909}.

\bibitem[{\textit{Clilverd et~al.}(2001)\textit{Clilverd, Rodger, Thomson, and
  Yearby}}]{Clilverd2001}
Clilverd, M.~A., C.~J. Rodger, N.~R. Thomson, and K.~H. Yearby (2001),
  {Investigating the possible association between thunderclouds and
  plasmaspheric ducts}, \textit{JGR}, \textit{106}(A12), 29,771 -- 29,781,
  \doi{10.1029/2001JA000081}.

\bibitem[{\textit{Cohen and R\"{o}ttgering}(2009)}]{Cohen2009}
Cohen, A.~S., and H.~J.~A. R\"{o}ttgering (2009), {Probing fine-scale
  ionospheric structure with the Very Large Array radio telescope},
  \textit{AJ}, \textit{138}(2), 439--447, \doi{10.1088/0004-6256/138/2/439}.

\bibitem[{\textit{Coker et~al.}(2009)\textit{Coker, Thonnard, Dymond, Lazio,
  Makela, and Loughmiller}}]{Coker2009}
Coker, C., S.~E. Thonnard, K.~F. Dymond, T.~J.~W. Lazio, J.~J. Makela, and
  P.~J. Loughmiller (2009), {Simultaneous radio interferometer and optical
  observations of ionospheric structure at the Very Large Array},
  \textit{RaSc}, \textit{44}(1), RS0A11, \doi{10.1029/2008RS004079}.

\bibitem[{\textit{Cole}(1971)}]{Cole1971}
Cole, K.~D. (1971), {Formation of field-aligned irregularities in the
  magnetosphere}, \textit{JATP}, \textit{33}, 741--750,
  \doi{10.1016/0021-9169(71)90027-4}.

\bibitem[{\textit{Condon et~al.}(1998)\textit{Condon, Cotton, Greisen, Yin,
  Perley, Taylor, and Broderick}}]{Condon1998}
Condon, J.~J., W.~D. Cotton, E.~W. Greisen, Q.~F. Yin, R.~A. Perley, G.~B.
  Taylor, and J.~J. Broderick (1998), {The NRAO VLA sky survey}, \textit{AJ},
  \textit{115}, 1693, \doi{10.1086/300337}.

\bibitem[{\textit{Coster et~al.}(2012)\textit{Coster, Herne, Erickson, and
  Oberoi}}]{Coster2012}
Coster, A., D.~Herne, P.~Erickson, and D.~Oberoi (2012), {Using the Murchison
  Widefield Array to observe midlatitude space weather}, \textit{RaSc},
  \textit{47}(6), RS0K07, \doi{10.1029/2012RS004993}.

\bibitem[{\textit{Dagg}(1957)}]{Dagg1957a}
Dagg, M. (1957), {The origin of the ionospheric irregularities responsible for
  radio-star scintillations and spread-F-II. Turbulent motion in the dynamo
  region}, \textit{Journal of Atmospheric and Terrestrial Physics},
  \textit{11}, 139 -- 150.

\bibitem[{\textit{Darrouzet et~al.}(2009)\textit{Darrouzet, Gallagher,
  Andr\'{e}, Carpenter, Dandouras, D\'{e}cr\'{e}au, {De Keyser}, Denton,
  Foster, Goldstein, Moldwin, Reinisch, Sandel, and Tu}}]{Darrouzet2009}
Darrouzet, F., D.~L. Gallagher, N.~Andr\'{e}, D.~L. Carpenter, I.~Dandouras,
  P.~M.~E. D\'{e}cr\'{e}au, J.~{De Keyser}, R.~E. Denton, J.~C. Foster,
  J.~Goldstein, M.~B. Moldwin, B.~W. Reinisch, B.~R. Sandel, and J.~Tu (2009),
  {Plasmaspheric density structures and dynamics: Properties observed by the
  CLUSTER and IMAGE missions}, \textit{Space Sci.~Rev.}, \textit{145}(1-2),
  55--106, \doi{10.1007/s11214-008-9438-9}.

\bibitem[{\textit{Dymond et~al.}(2011)}]{Dymond2011_mn2e}
Dymond, K.~F., et~al. (2011), {A medium-scale traveling ionospheric disturbance
  observed from the ground and from space}, \textit{RaSc}, \textit{46}(5),
  RS5010, \doi{10.1029/2010RS004535}.

\bibitem[{\textit{Fejer and Kelley}(1980)}]{Fejer1980}
Fejer, B.~G., and M.~C. Kelley (1980), {Ionospheric irregularities},
  \textit{RvGSP}, \textit{18}(2), 401, \doi{10.1029/RG018i002p00401}.

\bibitem[{\textit{Georges}(1968)}]{Georges1968}
Georges, T.~M. (1968), {HF Doppler studies of traveling ionospheric
  disturbances}, \textit{JATP}, \textit{30}(5), 735--746,
  \doi{10.1016/S0021-9169(68)80029-7}.

\bibitem[{\textit{Georges}(1973)}]{Georges1973}
Georges, T.~M. (1973), {Infrasound from convective storms: Examining the
  evidence}, \textit{RvGSP}, \textit{11}(3), 571--594,
  \doi{10.1029/RG011i003p00571}.

\bibitem[{\textit{Gold}(1959)}]{Gold1959}
Gold, T. (1959), {Motions in the magnetosphere of the Earth}, \textit{JGR},
  \textit{64}, 1219--1224, \doi{10.1029/JZ064i009p01219}.

\bibitem[{\textit{Hancock et~al.}(2012)\textit{Hancock, Murphy, Gaensler,
  Hopkins, and Curran}}]{Hancock2012}
Hancock, P.~J., T.~Murphy, B.~M. Gaensler, A.~Hopkins, and J.~R. Curran (2012),
  {Compact continuum source finding for next generation radio surveys},
  \textit{MNRAS}, \textit{422}(2), 1812--1824,
  \doi{10.1111/j.1365-2966.2012.20768.x}.

\bibitem[{\textit{Hawarey}(2006)}]{Hawarey2006}
Hawarey, M. (2006), {Travelling ionospheric disturbance over California mid
  2000}, \textit{Nonlin.~Proc.~Geoph.}, \textit{13}(1), 1--7,
  \doi{10.5194/npg-13-1-2006}.

\bibitem[{\textit{Helmboldt}(2012)}]{Helmboldt2012d}
Helmboldt, J.~F. (2012), {Insights into the nature of northwest-to-southeast
  aligned ionospheric wavefronts from contemporaneous Very Large Array and
  ionosonde observations}, \textit{JGR}, \textit{117}(A7), A07,310,
  \doi{10.1029/2012JA017802}.

\bibitem[{\textit{Helmboldt}(2014)}]{Helmboldt2014a}
Helmboldt, J.~F. (2014), {Drift-scan imaging of traveling ionospheric
  disturbances with the Very Large Array}, \textit{GRL}, \textit{41}, 1--9,
  \doi{10.1002/2014GL060951}.

\bibitem[{\textit{Helmboldt and Intema}(2012)}]{Helmboldt2012b}
Helmboldt, J.~F., and H.~T. Intema (2012), {Very large array observations of
  disturbed ion flow from the plasmasphere to the nighttime ionosphere},
  \textit{RaSc}, \textit{47}(6), RS0K03, \doi{10.1029/2012RS004979}.

\bibitem[{\textit{Helmboldt and Intema}(2014)}]{Helmboldt2014}
Helmboldt, J.~F., and H.~T. Intema (2014), {Advanced spectral analysis of
  ionospheric waves observed with sparse arrays}, \textit{JGRA}, \textit{119},
  1392--1413, \doi{10.1002/2013JA019162}.

\bibitem[{\textit{Helmboldt et~al.}(2012{\natexlab{a}})\textit{Helmboldt, Lane,
  and Cotton}}]{Helmboldt2012c}
Helmboldt, J.~F., W.~M. Lane, and W.~D. Cotton (2012{\natexlab{a}}),
  {Climatology of midlatitude ionospheric disturbances from the Very Large
  Array Low-frequency Sky Survey}, \textit{RaSc}, \textit{47}(5), RS5008,
  \doi{10.1029/2012RS005025}.

\bibitem[{\textit{Helmboldt et~al.}(2012{\natexlab{b}})\textit{Helmboldt,
  Lazio, Intema, and Dymond}}]{Helmboldt2012}
Helmboldt, J.~F., T.~J.~W. Lazio, H.~T. Intema, and K.~F. Dymond
  (2012{\natexlab{b}}), {A new technique for spectral analysis of ionospheric
  TEC fluctuations observed with the Very Large Array VHF system: From QP
  echoes to MSTIDs}, \textit{RaSc}, \textit{47}(4), RS0L02,
  \doi{10.1029/2011RS004787}.

\bibitem[{\textit{Hewish}(1951)}]{Hewish1951}
Hewish, A. (1951), The diffraction of radio waves in passing through a
  phase-changing ionosphere, \textit{RSPSA}, \textit{209}(1096), 81--96,
  \doi{10.1098/rspa.1951.0189}.

\bibitem[{\textit{Hines}(1960)}]{Hines1960}
Hines, C.~O. (1960), {Internal atmospheric gravity waves at ionospheric
  heights}, \textit{CaJPh}, \textit{38}(11), 1441--1481, \doi{10.1139/p60-150}.

\bibitem[{\textit{Hoogeveen and Jacobson}(1997{\natexlab{a}})}]{Hoogeveen1997a}
Hoogeveen, G.~W., and A.~R. Jacobson (1997{\natexlab{a}}), {Radio
  interferometer measurements of plasmasphere density structures during
  geomagnetic storms}, \textit{JGR}, \textit{102}(A7), 14,177 -- 14,188,
  \doi{10.1029/97JA00484}.

\bibitem[{\textit{Hoogeveen and Jacobson}(1997{\natexlab{b}})}]{Hoogeveen1997}
Hoogeveen, G.~W., and A.~R. Jacobson (1997{\natexlab{b}}), {Improved analysis
  of plasmasphere motion using the VLA radio interferometer}, \textit{AnGeo},
  \textit{15}, 236--245, \doi{10.1007/s00585-997-0236-6}.

\bibitem[{\textit{Hunsucker}(1982)}]{Hunsucker1982}
Hunsucker, R.~D. (1982), {Atmospheric gravity waves generated in the
  high-latitude ionosphere: A review}, \textit{RvGSP}, \textit{20}(2), 293,
  \doi{10.1029/RG020i002p00293}.

\bibitem[{\textit{Hunsucker}(1991)}]{Hunsucker1991}
Hunsucker, R.~D. (1991), \textit{{Radio Techniques for Probing the Terrestrial
  Ionosphere}}, Springer-Verlag.

\bibitem[{\textit{Hurley-Walker et~al.}(2014)}]{Hurley-Walker2014}
Hurley-Walker, N., et~al. (2014), {The Murchison Widefield Array Commissioning 
  Survey: A low-frequency catalogue of 14,110 compact radio sources over 6,100 
  square degrees}, \textit{PASA}, \textit{31}, e045, \doi{10.1017/pasa.2014.40}.

\bibitem[{\textit{Jacobson and Erickson}(1992{\natexlab{a}})}]{Jacobson1992}
Jacobson, A.~R., and W.~C. Erickson (1992{\natexlab{a}}), {Wavenumber-resolved
  observations of ionospheric waves using the Very Large Array radiotelescope},
  \textit{P\&SS}, \textit{40}(4), 447--455, \doi{10.1016/0032-0633(92)90163-I}.

\bibitem[{\textit{Jacobson and Erickson}(1992{\natexlab{b}})}]{Jacobson1992a}
Jacobson, A.~R., and W.~C. Erickson (1992{\natexlab{b}}), {A method for
  characterizing transient ionospheric disturbances using a large
  radiotelescope array}, \textit{A\&A}, \textit{257}, 401--409.

\bibitem[{\textit{Jacobson and Erickson}(1993)}]{Jacobson1993}
Jacobson, A.~R., and W.~C. Erickson (1993), Observations of electron density
  irregularities in the plasmasphere using the {VLA} radio interferometer,
  \textit{AnGeo}, \textit{11}(10), 869 -- 888.

\bibitem[{\textit{James}(2006)}]{James2006}
James, H.~G. (2006), {Characteristics of field-aligned density depletion
  irregularities in the auroral ionosphere that duct Z- and X-mode waves},
  \textit{JGR}, \textit{111}(A9), A09,315, \doi{10.1029/2006JA011652}.

\bibitem[{\textit{Kazimirovsky et~al.}(2003)\textit{Kazimirovsky, Herraiz, and
  de~la Morena}}]{Kazimirovsky2002}
Kazimirovsky, E., M.~Herraiz, and B.~A. de~la Morena (2003), {Effects on the
  ionosphere due to phenomena occurring below it}, \textit{Surv.~Geophys.},
  \textit{24}, 139--184, \doi{10.1023/A:1023206426746}.

\bibitem[{\textit{Kulkarni et~al.}(2008)\textit{Kulkarni, Inan, Bell, and
  Bortnik}}]{Kulkarni2008}
Kulkarni, P., U.~S. Inan, T.~F. Bell, and J.~Bortnik (2008), {Precipitation
  signatures of ground-based VLF transmitters}, \textit{JGR}, \textit{113}(A7),
  A07,214, \doi{10.1029/2007JA012569}.

\bibitem[{\textit{Lester and Smith}(1980)}]{Lester1980}
Lester, M., and A.~J. Smith (1980), {Whistler duct structure and formation},
  \textit{Planetary and Space Science}, \textit{28}, 645--654,
  \doi{10.1016/0032-0633(80)90011-2}.

\bibitem[{\textit{Liu et~al.}(2011)\textit{Liu, Chen, Lin, Tsai, Chen, and
  Kamogawa}}]{Liu2011}
Liu, J.-Y., C.-H. Chen, C.-H. Lin, H.-F. Tsai, C.-H. Chen, and M.~Kamogawa
  (2011), {Ionospheric disturbances triggered by the 11 March 2011 M 9.0 Tohoku
  earthquake}, \textit{JGR}, \textit{116}(A6), A06,319,
  \doi{10.1029/2011JA016761}.

\bibitem[{\textit{Lloyd et~al.}(1972)\textit{Lloyd, Low, McAvaney, Rees, and
  Roper}}]{Lloyd1972}
Lloyd, K.~H., C.~H. Low, B.~J. McAvaney, D.~Rees, and R.~G. Roper (1972),
  Thermospheric observations combining chemical seeding and ground-based
  techniques -- i. winds, turbulence and the parameters of the neutral
  atmosphere, \textit{Planetary and Space Science}, \textit{20}, 76--789,
  \doi{10.1016/0032-0633(72)90159-6}.

\bibitem[{\textit{Loi et~al.}(2015{\natexlab{a}})}]{Loi2015_mn2e}
Loi, S.~T., et~al. (2015{\natexlab{a}}), Real-time imaging of density ducts
  between the plasmasphere and ionosphere, \textit{GRL}, \textit{42}(10),
  3707--3714, \doi{10.1002/2015GL063699}.

\bibitem[{\textit{Loi et~al.}(2015{\natexlab{b}})}]{Loi2015a_mn2e}
Loi, S.~T., et~al. (2015{\natexlab{b}}), {Power spectrum analysis of
  ionospheric fluctuations with the Murchison Widefield Array}, \textit{Radio
  Science}, \textit{50}, 574--597, \doi{10.1002/2015RS005711}.

\bibitem[{\textit{Loi et~al.}(2015{\natexlab{c}})}]{Loi2015b_mn2e}
Loi, S.~T., et~al. (2015{\natexlab{c}}), {Quantifying ionospheric effects on
  time-domain astrophysics with the Murchison Widefield Array}, \textit{MNRAS},
  \textit{453}(3), 2731--2746, \doi{10.1093/mnras/stv1808}.

\bibitem[{\textit{Lonsdale}(2005)}]{Lonsdale2005}
Lonsdale, C.~J. (2005), Configuration considerations for low frequency arrays,
  in \textit{From Clark Lake to the Long Wavelength Array: Bill Erickson's
  Radio Science}, vol. 345, edited by K.~E. Kassim, M.~R. P\'{e}rez, W.~Junor,
  and P.~A. Henning, p. 399, ASP Conf.~Ser.

\bibitem[{\textit{Lonsdale et~al.}(2009)}]{Lonsdale2009_mn2e}
Lonsdale, C.~J., et~al. (2009), {The Murchison Widefield Array: Design
  Overview}, \textit{Proc.~IEEE}, \textit{97}(8), 1497--1506,
  \doi{10.1109/JPROC.2009.2017564}.

\bibitem[{\textit{Mannucci et~al.}(1998)\textit{Mannucci, Wilson, Yuan, Ho,
  Lindqwister, and Runge}}]{Mannucci1998}
Mannucci, A.~J., B.~D. Wilson, D.~N. Yuan, C.~H. Ho, U.~J. Lindqwister, and
  T.~F. Runge (1998), {A global mapping technique for GPS derived ionospheric
  total electron content measurements}, \textit{RaSc}, \textit{33}(3),
  565--582, \doi{10.1029/97RS02707}.

\bibitem[{\textit{Mauch et~al.}(2003)\textit{Mauch, Murphy, Buttery, Curran,
  Hunstead, Robertson, and Sadler}}]{Mauch2003}
Mauch, T., T.~Murphy, H.~J. Buttery, J.~Curran, R.~W. Hunstead, J.~G.
  Robertson, and E.~M. Sadler (2003), {SUMSS: A wide-field radio imaging survey
  of the southern sky II. The source catalogue}, \textit{MNRAS}, \textit{342},
  1117--1130, \doi{10.1046/j.1365-8711.2003.06605.x}.

\bibitem[{\textit{McCormick}(2002)}]{McCormick2002}
McCormick, R.~J. (2002), {Reconsidering the effectiveness of quasi-static
  thunderstorm electric fields for whistler duct formation}, \textit{JGR},
  \textit{107}(A11), 1396, \doi{10.1029/2001JA009219}.

\bibitem[{\textit{McIlwain}(1961)}]{McIlwain1961}
McIlwain, C.~E. (1961), {Coordinates for mapping the distribution of
  magnetically trapped particles}, \textit{JGR}, \textit{66}(11), 3681--3691,
  \doi{10.1029/JZ066i011p03681}.

\bibitem[{\textit{Mercier}(1986)}]{Mercier1986}
Mercier, C. (1986), {Observations of atmospheric gravity waves by
  radiointerferometry}, \textit{JATP}, \textit{48}(7), 605--624,
  \doi{10.1016/0021-9169(86)90010-3}.

\bibitem[{\textit{Meyer-Vernet}(1980)}]{Meyer-Vernet1980}
Meyer-Vernet, N. (1980), On a day-time ionospheric effect on some radio
  intensity measurements and interferometry, \textit{A\&A}, \textit{84},
  142--147.

\bibitem[{\textit{Narayan and Goodman}(1989)}]{Narayan1989}
Narayan, R., and J.~Goodman (1989), {The shape of a scatter-broadened image -
  I. Numerical simulations and physical principles}, \textit{MNRAS},
  \textit{238}, 963 -- 994.

\bibitem[{\textit{Newcomb}(1961)}]{Newcomb1961}
Newcomb, W.~A. (1961), Convective instability induced by gravity in a plasma
  with a frozen-in magnetic field, \textit{Physics of Fluids}, \textit{4},
  391--396, \doi{10.1063/1.1706342}.

\bibitem[{\textit{Offringa et~al.}(2015)}]{Offringa2015}
Offringa, A.~O., et~al. (2015), The low-frequency environment of the murchison
  widefield array: Radio-frequency interference analysis and mitigation,
  \textit{PASA}, \textit{32}, e8, \doi{10.1017/pasa.2015.7}.

\bibitem[{\textit{Offringa et~al.}(2014)\textit{Offringa, McKinley,
  Hurley-Walker et~al.}}]{Offringa2014}
Offringa, A.~R., B.~McKinley, N.~Hurley-Walker, et~al. (2014), Wsclean: an
  implementation of a fast, generic wide-field imager for radio astronomy,
  \textit{MNRAS}, \textit{444}(1), 606--619, \doi{10.1093/mnras/stu1368}.

\bibitem[{\textit{Park}(1970)}]{Park1970}
Park, C.~G. (1970), Whistler observations of the interchange of ionization
  between the ionosphere and the protonosphere, \textit{JGR}, \textit{75}(22),
  4249 -- 4260, \doi{10.1029/JA075i022p04249}.

\bibitem[{\textit{Park}(1971)}]{Park1971}
Park, C.~G. (1971), {Westward electric fields as the cause of nighttime
  enhancements in electron concentrations in midlatitude F region},
  \textit{JGR}, \textit{76}(19), 4560--4568, \doi{10.1029/JA076i019p04560}.

\bibitem[{\textit{Park and Dejnakarintra}(1973)}]{Park1973}
Park, C.~G., and M.~Dejnakarintra (1973), {Penetration of thundercloud electric
  fields into the ionosphere and magnetosphere: 1. Middle and subauroral
  latitudes}, \textit{JGR}, \textit{78}(28), 6623 -- 6633,
  \doi{10.1029/JA078i028p06623}.

\bibitem[{\textit{Platt and Dyson}(1989)}]{Platt1989}
Platt, I., and P.~Dyson (1989), {MF and HF propagation characteristics of
  ionospheric ducts}, \textit{JATP}, \textit{51}, 759--774,
  \doi{10.1016/0021-9169(89)90033-0}.

\bibitem[{\textit{Reid}(1965)}]{Reid1965}
Reid, G.~C. (1965), Ionospheric effects of electrostatic fields generated in
  the outer magnetosphere, \textit{Radio Science}, \textit{69D}(6), 827--837.

\bibitem[{\textit{Richmond}(1978)}]{Richmond1978}
Richmond, A.~D. (1978), {Gravity wave generation, propagation, and dissipation
  in the thermosphere}, \textit{JGR}, \textit{83}(A9), 4131--4145,
  \doi{10.1029/JA083iA09p04131}.

\bibitem[{\textit{Rideout and Coster}(2006)}]{Rideout2006}
Rideout, W., and A.~Coster (2006), {Automated GPS processing for global total
  electron content data}, \textit{GPS Solut.}, \textit{10}(3), 219--228,
  \doi{10.1007/s10291-006-0029-5}.

\bibitem[{\textit{Rodger et~al.}(2002)\textit{Rodger, Thomson, and
  Dowden}}]{Rodger2002}
Rodger, C.~J., N.~R. Thomson, and R.~L. Dowden (2002), Correction to "are
  whistler ducts created by thunderstorm electric fields?" by c.~j.~rodger et
  al., \textit{Journal of Geophysical Research}, \textit{107}, 1068,
  \doi{10.1029/2001JA009152}.

\bibitem[{\textit{Rodger et~al.}(1998)\textit{Rodger, Thomson, and
  Dowden}}]{Rodger1998}
Rodger, J., N.~R. Thomson, and L.~Dowden (1998), {Are whistler ducts created by
  thunderstorm electrostatic fields?}, \textit{JGR}, \textit{103}(A2),
  2163--2169, \doi{10.1029/97JA02927}.

\bibitem[{\textit{Rowlinson et~al.}(2015, submitted)}]{Rowlinson2015_mn2e}
Rowlinson, A., et~al. (2015, submitted), {Limits on FRBs and other transient
  sources at 182 MHz using the Murchison Widefield Array}, \textit{MNRAS}.

\bibitem[{\textit{Sagredo and Bullough}(1973)}]{Sagredo1973}
Sagredo, J.~L., and K.~Bullough (1973), {VLF goniometer observations at Halley
  Bay, Antarctica - II. Magnetospheric structure deduced from whistler
  observations}, \textit{Planetary and Space Science}, \textit{21}, 913 -- 923,
  \doi{10.1016/0032-0633(73)90139-6}.

\bibitem[{\textit{Singh et~al.}(1994)\textit{Singh, Singh, Singh, and
  Singh}}]{Singh1994}
Singh, R.~P., A.~K. Singh, U.~P. Singh, and R.~N. Singh (1994), {Wave ducting
  and scattering properties of ionospheric irregularities}, \textit{Advances in
  Space Research}, \textit{14}(9), 225--228,
  \doi{10.1016/0273-1177(94)90140-6}.

\bibitem[{\textit{Singh et~al.}(1998)\textit{Singh, Singh, and
  Singh}}]{Singh1998}
Singh, R.~P., A.~K. Singh, and D.~K. Singh (1998), {Plasmaspheric parameters as
  determined from whistler spectrograms: A review},
  \textit{J.~Atmos.~Sol.-Terr.~Phys.}, \textit{60}(5), 495--508,
  \doi{10.1016/S1364-6826(98)00001-7}.

\bibitem[{\textit{Smith and Clilverd}(1991)}]{Smith1991}
Smith, A.~J., and M.~A. Clilverd (1991), {Magnetic storm effects on the
  mid-latitude plasmasphere}, \textit{Planetary and Space Science},
  \textit{39}(7), 1069--1079, \doi{10.1016/0032-0633(91)90114-P}.

\bibitem[{\textit{Smith}(1952)}]{Smith1952}
Smith, F.~G. (1952), {Ionospheric refraction of 81.5 Mc/s radio waves from
  radio stars}, \textit{JATP}, \textit{2}(August), 350--355.

\bibitem[{\textit{Smith et~al.}(1960)\textit{Smith, Helliwell, and
  Yabroff}}]{Smith1960}
Smith, R.~L., R.~A. Helliwell, and I.~W. Yabroff (1960), A theory of trapping
  of whistlers in field-aligned columns of ionization, \textit{Journal of
  Geophysical Research}, \textit{65}(3), 815 -- 823.

\bibitem[{\textit{Sonwalkar}(2006)}]{Sonwalkar2006}
Sonwalkar, V.~S. (2006), The influence of plasma density irregularities on
  whistler-mode wave propagation, in \textit{Geospace electromagnetic waves and
  radiation}, vol. 191, pp. 141--190, Springer, \doi{10.1007/3-540-33203-0\_6}.

\bibitem[{\textit{Sonwalkar et~al.}(1994)\textit{Sonwalkar, Inan, Bell,
  Helliwell, Chmyrev, Sobolev, Ovcharenko, and Selegej}}]{Sonwalkar1994}
Sonwalkar, V.~S., U.~S. Inan, T.~F. Bell, R.~A. Helliwell, V.~M. Chmyrev, Y.~P.
  Sobolev, O.~Y. Ovcharenko, and V.~Selegej (1994), {Simultaneous observations
  of VLF ground transmitter signals on the DE 1 and COSMOS 1809 satellites:
  Detection of a magnetospheric caustic and a duct}, \textit{JGR},
  \textit{99}(A9), 17,511--17,522, \doi{10.1029/94JA00866}.

\bibitem[{\textit{Spoelstra}(1985)}]{Spoelstra1985}
Spoelstra, T. A.~T. (1985), Effects of amplitude and phase scintillations on
  decimeter wavelength observations at mid-latitudes, \textit{A\&A},
  \textit{148}, 21 -- 28.

\bibitem[{\textit{Spoelstra}(1997)}]{Spoelstra1997}
Spoelstra, T. A.~T. (1997), {The ionosphere and radio interferometry},
  \textit{AnGeo}, \textit{XL}(4), 865 -- 885, \doi{10.4401/ag-3885}.

\bibitem[{\textit{Storey}(1953)}]{Storey1953}
Storey, L. R.~O. (1953), {An investigation of whistling atmospherics},
  \textit{Philosophical Transactions of the Royal Society of London Series A,
  Mathematical and Physical Sciences}, \textit{246}(908), 113--141,
  \doi{10.1098/rsta.1953.0011}.

\bibitem[{\textit{Sugiura}(1964)}]{Sugiura1964}
Sugiura, M. (1964), \textit{Hourly values of equatorial Dst for the IGY},
  vol.~35, 9--45 pp., Pergamon Press, Oxford.

\bibitem[{\textit{Th\'{e}bault et~al.}(2015)}]{Thebault2015}
Th\'{e}bault, E., et~al. (2015), International geomagnetic reference field: the
  12th generation, \textit{Earth, Planets and Space}, \textit{67}, 79,
  \doi{10.1186/s40623-015-0228-9}.

\bibitem[{\textit{Tingay et~al.}(2013)}]{Tingay2013_mn2e}
Tingay, S.~J., et~al. (2013), {The Murchison Widefield Array: The Square
  Kilometre Array precursor at low radio frequencies}, \textit{PASA},
  \textit{30}, e007, \doi{10.1017/pasa.2012.007}.

\bibitem[{\textit{van Haarlem et~al.}(2013)\textit{van Haarlem, Wise, Gunst
  et~al.}}]{vanHaarlem2013}
van Haarlem, M.~P., M.~W. Wise, A.~W. Gunst, et~al. (2013), Lofar: The
  low-frequency array, \textit{A\&A}, \textit{556}, A2,
  \doi{10.1051/0004-6361/201220873}.

\bibitem[{\textit{Voss et~al.}(1984)\textit{Voss, Imhof, Walt, Mobilia, Gaines,
  Reagan, Inan, Helliwell, Carpenter, Katsufrakis, and Chang}}]{Voss1984}
Voss, H.~D., W.~L. Imhof, M.~Walt, J.~Mobilia, E.~E. Gaines, J.~B. Reagan,
  U.~S. Inan, R.~A. Helliwell, D.~L. Carpenter, J.~P. Katsufrakis, and H.~C.
  Chang (1984), {Lightning-induced electron precipitation}, \textit{Nature},
  \textit{312}, 740--742, \doi{10.1038/312740a0}.

\bibitem[{\textit{Walker}(1978)}]{Walker1978}
Walker, A. D.~M. (1978), {Formation of whistler ducts}, \textit{P\&SS},
  \textit{26}, 375 -- 379, \doi{10.1051/0004-6361/201220873}.

\bibitem[{\textit{Weber et~al.}(1978)\textit{Weber, Buchau, Eather, and
  Mende}}]{Weber1978}
Weber, E.~J., J.~Buchau, R.~H. Eather, and S.~B. Mende (1978), {North-south
  aligned equatorial airglow depletions}, \textit{JGR}, \textit{83}(A2), 712,
  \doi{10.1029/JA083iA02p00712}.

\bibitem[{\textit{Wild and Roberts}(1956)}]{Wild1956}
Wild, J.~P., and J.~A. Roberts (1956), {The spectrum of radio-star
  scintillations and the nature of irregularities in the ionosphere},
  \textit{Journal of Atmospheric and Terrestrial Physics}, \textit{8}, 55 --
  75, \doi{10.1016/0021-9169(56)90091-5}.

\end{thebibliography}
\end{document}